\documentstyle[11pt]{article}

\newtheorem{teor}{Theorem}
\newtheorem{prop}{Proposition}
\newtheorem{corol}{Corollary}
\newtheorem{lem}{Lemma}
\newtheorem{definition}{Definition}

\newtheorem{assum}{Assumption}

\def\beq{\begin{equation}}
\def\eeq{\end{equation}}
\def\bea{\begin{eqnarray}}
\def\eea{\end{eqnarray}}
\def\beann{\begin{eqnarray*}}
\def\eeann{\end{eqnarray*}}
\def\beasn{\begin{sneqnarray}}
\def\eeasn{\end{sneqnarray}}
\def\ben{\begin{enumerate}}
\def\een{\end{enumerate}}
\def\bit{\begin{itemize}}
\def\eit{\end{itemize}}
\def\dst{\(\displaystyle}
\def\proof{( {\sl Proof} )\quad}

\def\derpar#1#2{\frac{\partial{#1}}{\partial{#2}}}

\def\mapping#1{\mathrel{\mathop{\longrightarrow}\limits^{#1}}}

\def\forta#1{\mathop{\mathpalette\@vereq\sim}\limits_{#1}}
\def\coor#1#2#3{{#1}^{#2}, \ldots, {#1}^{#3}}
\def\moment#1#2#3{{#1}_{#2}, \ldots, {#1}_{#3}}

\def\qed{\ifvmode\removelastskip\fi
{\unskip\nobreak\hfil\penalty50\hbox{}\nobreak\hfil
\hbox{\vrule height1.2ex width1.2ex}\parfillskip=0pt
\finalhyphendemerits=0 \par\smallskip}}

\def\vf{{\cal X}}
\def\df{{\mit\Omega}}
\def\Lag{{\cal L}}

\def\d{{\rm d}}

\def\Real{{\bf R}}

\def\inn{\mathop{i}\nolimits}

\def\Tan{{\rm T}}

\def\Lie{\mathop{\rm L}\nolimits}
\def\Cinfty{{\rm C}^\infty}

\catcode`@=12

\def\tabaddress#1{{\it\begin{tabular}[t]{c}#1 \\[1.2ex]\end{tabular}}}
\def\UPCMAT{Departamento de Matem\'atica Aplicada y Telem\'atica\\
   Campus Norte U.P.C., M\'odulo C-3\\
   C/ Jordi Girona 1\\
   E-08034 Barcelona, Spain}

\begin{document}

\title{REDUCTION OF PRESYMPLECTIC MANIFOLDS WITH SYMMETRY}
\author{\sc A. Echeverr\'ia-Enr\'iquez,
M. C. Mu\~noz-Lecanda\thanks{{\bf e}-{\it mail}: MATMCML@MAT.UPC.ES},
N. Rom\'an-Roy\thanks{{\bf e}-{\it mail}: MATNRR@MAT.UPC.ES}
   \\
   \tabaddress{\UPCMAT}}
\date{To be published in {\sl Rev. Math. Phys.} {\bf 11}(10) 1999}

\pagestyle{myheadings}
\markright{\sc A. Echeverr\'\i a {\it et al}:
{\sl Reduction of presymplectic manifolds with symmetry.}}

\maketitle
\thispagestyle{empty}
\setcounter{page}{0}

\begin{abstract}
Actions of Lie groups on presymplectic manifolds are analyzed,
introducing the suitable comomentum and momentum maps.
The subsequent theory of reduction of
presymplectic dynamical systems with symmetry is studied.
In this way, we give a method of reduction which
enables us to remove gauge symmetries as well as non-gauge
``rigid'' symmetries at once.
This method is compared with other step-by-step reduction procedures.
As particular examples in this framework, we discuss
the reduction of time-dependent dynamical systems with symmetry,
the reduction of a mechanical model of field theories
with gauge and non-gauge symmetries, and the gauge reduction of the system
made of a conformal particle.
\end{abstract}

\vfill \hfill
\vbox{\raggedleft AMS s.\,c.\,(1991): 57S25, 58D19, 70H33.
 PACS: 0240, 0320 }\null

\clearpage

\section{Introduction}

The problem of reduction of dynamical systems with symmetry
has deserved the interest of theoretical physicists
and mathematicians, with the purpose of reducing the number of
evolution equations, by finding first integrals of motion.
In particular, geometric treatment of this subject
has been revealed as a powerful tool in the study of this question.
The pioneering and fundamental work on this topic
has been carried out by Marsden and Weinstein
\cite{MW-rsms} (see also \cite{AM-78}, \cite{LM-87} and \cite{We-lsm}).
They demonstrated that, for a free and proper symplectic action
of a (connected) Lie group on a (connected) symplectic manifold
(which is the phase space of an {\sl autonomous regular Hamiltonian system}
with symmetry), and a {\sl weakly regular value} of the {\sl momentum map}
associated with this action, the {\sl reduced phase space}
has a structure of symplectic manifold and inherits a Hamiltonian
dynamics from the initial system.

Nevertheless, the problem of reduction can appear under
many different aspects. Subsequently, other authors have investigated
aspects of the theory of reduction for other particular cases.

Thus, for instance, if zero is a {\sl singular value}
of the momentum map (in a symplectic manifold)
then the Marsden-Weinstein technique
gives a reduced phase space which is a {\sl stratified symplectic space}
\cite{SL-91}. Starting from this result, reduction of
{\sl time-dependent regular Hamiltonian systems} with
momentum mappings with singular value at zero is achieved in \cite{LS-93},
where, using the {\sl extended phase space symplectic formalism},
it is proved that the reduced phase space is also a {\sl stratified space}
but with a cosymplectic structure.
Another approach to the problem of singular values
can be found in \cite{ACG-91} (see also other references quoted therein),
where reduction of symplectic manifolds
at singular values of the momentum mapping is considered, showing that,
under certain conditions, the reduced space inherits a
non-degenerate Poisson structure.
However, research in this area is not yet complete.

In the realm of momentum maps with {\sl regular values},
the Marsden-Weinstein symplectic reduction scheme
has been applied to many different situations.
For example, reduction of
{\sl time-dependent regular Hamiltonian systems}
is developed in the framework of cosymplectic manifolds in \cite{Al-89},
obtaining a reduced phase space which inherits
a structure of cosymplectic manifold.
The study of {\sl autonomous singular Lagrangian systems}
can be found in \cite{CCCI-86} and, in particular, the conditions for
the reduced phase space to inherit an almost-tangent structure
are studied for certain kinds of degenerate Lagrangians.
Some of the results here obtained are generalized
to the case of {\sl non-autonomous singular Lagrangian systems}
and for a larger class of degenerate Lagrangians in \cite{IM-92}.
Another approach to this question is made in \cite{LMR-92}, where
the authors analyze the conditions for the existence of a regular
Lagrangian function in the reduced phase space obtained after reduction,
in such a way that the reduced cosymplectic or contact structure
(and hence the reduced Hamiltonian function) can be constructed from it.

Furthermore, there are other situations in reduction theory. So, for instance,
the theory of reduction of {\sl Poisson manifolds}
is treated in works such as \cite{LM-95} and \cite{MR-86}.
Reduction of cotangent bundles of {\sl Lie groups}
within {\sl semidirect products}
is considered in \cite{MRW-84}, with several applications
to outstanding problems in mathematical physics.
Concerning the subject of {\sl Lagrangian reduction},
there are some works, such as \cite{MS-93},
which consider the problem from the point of view of
reducing variational principles (instead of reducing the
almost tangent structure, as it is
made in some of the above mentioned references).
Finally, the study of reduction of {\sl non-holonomic systems}
can be found, for instance, in \cite{BS-93}, \cite{CLMM-98} and \cite{Ma-95}.
(Of course, this list of references is far to be complete).

The aim of this work is to apply the Marsden-Weinstein method to reduce
{\sl presymplectic manifolds} with Lie groups of symmetries acting on them.
The interest of this topic lies in the fact that
the geometrical description of many dynamical systems
is given by means of presymplectic manifolds.
One of the more frequent cases is the
{\sl Lagrangian formalism of singular mechanical systems},
where the phase space is the manifold $\Tan Q$
($Q$ being the configuration manifold of the system),
endowed with the presymplectic form $\Omega_{\Lag}$,
which is constructed from the singular Lagrangian function $\Lag$.
Other typical examples are certain descriptions of
non-autonomous mechanical systems
(both in the Lagrangian and Hamiltonian formalism),
where the phase space is a {\sl contact} ({\sl cosymplectic}) manifold.
Certainly, these kind of systems could be reduced by
first constructing an ambient symplectic manifold
where the system is coisotropically imbedded,
and then applying the symplectic reduction procedure to it \cite{IM-95}.
But we give a reduction procedure that allows us
to implement the Marsden-Weinstein technique directly for the
initial presymplectic system.

In particular, we construct comomentum and momentum mappings for
presymplectic actions of Lie groups, analyzing the obstruction to
their existence and studying some characteristics features
of the level sets of the momentum map.
Then, we prove that, for weakly regular values
of this momentum map, and under the usual suitable assumptions,
the reduced phase space inherits a presymplectic structure.
Next we apply these results in order to reduce
{\sl presymplectic dynamical systems with symmetry},
showing that, if we consider together {\sl gauge} and {\sl non-gauge} (``rigid'') symmetries, and we reduce the system by all of them,
then this procedure leads to the same results as if
we first remove the gauge redundancy and then reduce
the remaining ``rigid'' symmetries.
Finally, we analyze three examples,
namely: non-autonomous dynamical systems with symmetry
(comparing then the results so obtained with those of some
of the above mentioned references),
a mechanical model for field theories,
and the {\sl conformal particle}.

The paper is organized in the following way:

The first part is devoted to the study of
{\sl presymplectic group actions}. Thus, in sections \ref{ps} and \ref{algpm},
we review some basic concepts on presymplectic manifolds
and present the actions of Lie groups on them.
In sections \ref{mm} and \ref{lsmm}
we define the comomentum and momentum mappings
for this kind of actions, studying the obstruction to their existence,
the level sets of the momentum map and their reduction.

The second part deals with symmetries of presymplectic
dynamical systems. First, in section \ref{rpds},
we review the basic features of this kind of dynamical systems.
Section \ref{spds} is devoted to defining and analyzing
the concept of symmetry for these systems and to establish the
reduction procedure for compatible presymplectic systems.
The reduction procedure for non-compatible presymplectic systems and
its characteristic features is established in section \ref{rpdss}.
This part ends with a comparative study between this reduction
method and other different ways for reducing presymplectic systems,
which is performed in sections \ref{comp1} and \ref{coim}.

In the third part some examples are analyzed.
In sections \ref{rnass} and \ref{ads}
these techniques are applied in order to make
the reduction of non-autonomous systems with symmetry and,
as a particular example, the dynamics of autonomous regular dynamical systems
is obtained in this context.
A further example is the complete reduction of a
particular case of a mechanical model of field theories coupled to
external fields (due to {\it Capri} and {\it Kobayashi}),
which is investigated in sections \ref{mmft1} and \ref{mmft2}.
As the last example, the gauge reduction of the system of
a conformal particle is discussed, in this framework,
in section \ref{cp}.

Finally, we discuss the results and compare them with
those obtained in some of the works above mentioned.

An appendix is devoted to a linear interpretation
of the reduction theory.

All the manifolds are real, connected, second countable and $\Cinfty$.
The maps are assumed to be $\Cinfty$
and the differential forms have constant rank.
Sum over crossed repeated indices is understood.
We will denote by $\vf (M)$, $\df^p(M)$ and $\Cinfty (M)$
the sets of vector fields, differentiable $p$-forms and functions
in the manifold $M$ respectively. Finally
$\inn (X)\alpha$ will denote
the inner product or contraction of $X\in\vf (M)$ with $\alpha\in\df^p(M)$
and $\Lie (X)\alpha$ the Lie derivative of the form $\alpha$
along the vector field $X$.
Finally, along the work, quotient of manifolds by involutive distributions
will be made; and then we assume that the corresponding quotient spaces
are differentiable manifolds (conditions in order to assure this fact
are stated in \cite{AMR-83}).

\section{Presymplectic group actions}

\subsection{Presymplectic manifolds: previous statements}
\protect\label{ps}

Let us first recall that a {\sl presymplectic manifold}
is a couple $(M,\Omega )$ where $M$ is a $m$-dimensional
differentiable manifold and
$\Omega\in\df^2(M)$ is a closed degenerate differentiable form in $M$.
Let
$$
\ker\,\Omega :=\{ Z\in\vf (M)\ \vert\ \inn (Z)\Omega =0\}
$$
which is assumed to be a distribution on $M$ (that is, it has constant rank).

A vector field $X\in\vf (M)$
is said to be a {\sl Hamiltonian vector field}
(with respect to the presymplectic structure $\Omega$)
iff $\inn (X)\Omega$ is an exact 1-form; that is,
there exists $f_X\in\Cinfty (M)$ such that
\beq
\inn (X)\Omega =\d f_X
\label{ham}
\eeq
We will denote by $\vf_h(M)$ the set of Hamiltonian vector fields in $M$.

$X\in\vf (M)$ is said to be a
{\sl locally Hamiltonian vector field}
(with respect to the presymplectic structure $\Omega$)
iff $\inn (X)\Omega$ is a closed 1-form.
In this case, for every point $x\in M$, there is an open
neighbourhood $U\subset M$ and $f\in\Cinfty (U)$ such that
$$
\inn (X)\Omega\vert_U= \d f
$$
We will denote by $\vf_{lh}(M)$ the set of
locally Hamiltonian vector fields in $M$, and it is obvious that
$\vf_h(M)\subset\vf_{lh}(M)$.
On the other hand, it is also immediate to observe that
$X\in\vf_{lh}(M)$ if, and only if, $\Lie (X)\Omega =0$.
Finally, for every $X\in\vf_{lh}(M)$ and
$Z\in\ker\,\Omega$, we have that $[X,Z]\in\ker\,\Omega$.

$f\in\Cinfty (M)$ is said to be a {\sl presymplectic Hamiltonian function}
iff there exist a vector field $X\in\vf (M)$ such that (\ref{ham}) holds.
We will denote by $X_f$ the Hamiltonian vector field
associated with $f$ and by $\Cinfty_h (M)$ the set of
presymplectic Hamiltonian functions in $M$.
If $f$ is a presymplectic Hamiltonian function
then $\Lie(Z)f=0$, for every $Z\in\ker\,\Omega$
(and the same results holds for locally Hamiltonian functions
in $U\subset M$).

Since $\ker\,\Omega\subset\vf_h(M)$,
then, if $X\in\vf_h(M)$ and $Z\in\ker\,\Omega$,
then $f_X=f_{X+Z}$ and, conversely,
if $X,Y\in\vf_h(M)$ and $f_X=f_Y$, therefore a vector field
$Z\in\ker\,\Omega$ exists such that $X=Y+Z$.
On the other hand, if $f\in\Cinfty_h(M)$ and $\lambda\in\Real$
then $X_f=X_{f+\lambda}$ and, conversely, if
$f,g\in\Cinfty_h(M)$ and $X_f=X_g$ then there exists $\lambda\in\Real$
such that $f=g+\lambda$
(remember that $M$ is supposed to be connected).

Let $f_1,f_2\in\Cinfty_h(M)$ be presymplectic
Hamiltonian functions and $X_1,X_2\in\vf_h(M)$
Hamiltonian vector fields for these functions.
The {\sl Poisson bracket} of these Hamiltonian functions
(related to the presymplectic structure $\Omega$) is the
function $\{ f_1,f_2\}$ given by
$$
\{ f_1,f_2\}:=\Omega (X_1,X_2)=
\inn (X_2)\inn (X_1)\Omega= \inn (X_2)\d f_1=-\inn (X_1)\d f_2
$$
It is trivial to prove that this definition does not depend
on the Hamiltonian vector fields we have chosen.
In addition, $\{ f_1,f_2\}\in\Cinfty_h(M)$ and
$\inn ([X_1,X_2])\Omega =\d\{ f_2,f_1\}$, in fact,
$$
\inn ([X_1,X_2])\Omega=
\Lie (X_1)\inn (X_2)\Omega-\inn (X_1)\Lie (X_2)\Omega=
\Lie (X_1)\inn (X_2)\Omega =\Lie (X_1)\d f_2=
\d\{ f_2,f_1\}
$$
hence, $\vf_h(M)$ is a Lie subalgebra of $\vf (M)$.
The same thing holds for $\vf_{lh}(M)$ and $\ker\,\Omega$
is an ideal of both algebras.
So we have a map $(f_1,f_2)\mapsto\{ f_1,f_2\}$
defined in $\Cinfty_h(M)/\Real$ which transforms it into a real Lie algebra.
In order to prove the Jacobi identity, observe that, from the last equality, we obtain that
$\Omega ([X_1,X_2],X_3)=-\{ f_3,\{ f_2,f_1\}\}$.

Considering the map
$\Omega^\sharp\colon\vf (M)\to\df^1(M)$ defined by
$\Omega^\sharp (X):=\inn (X)\Omega$, for every $X\in\vf (M)$,
its restriction $\Omega^\sharp_h\colon\vf_h(M)\to\d\Cinfty_h(M)$
goes down to the quotient $\vf_{h}(M)/\ker\,\Omega$,
which is a Lie algebra because $\ker\,\Omega$ is an ideal of
the Lie algebra $\vf_h(M)$, and hence
the map
\dst\tilde\Omega\colon\vf_{h}(M)/\ker\,\Omega\to\Cinfty_h(M)/\Real\)
is bijective and, according to the previous remark,
a Lie algebra (anti) isomorphism.

\subsection{Actions of Lie groups on presymplectic manifolds}
\protect\label{algpm}

Let $G$ be a Lie group (which we will assume to be connected),
${\bf g}$ its Lie algebra, $(M,\Omega )$ a
presymplectic manifold and
$\Phi \colon G\times M \to M$ a presymplectic action of $G$ on $M$;
that is, $\Phi_g^*\Omega = \Omega$, for every $g\in G$.
As a consequence, the fundamental vector field $\tilde\xi\in{\cal X}(M)$,
associated with every $\xi\in{\bf g}$ by $\Phi$, is a
locally Hamiltonian vector field, $\tilde\xi\in {\cal X}_{lh}(M)$
(conversely, if for every $\xi\in{\bf g}$, we have that
$\tilde\xi\in\vf_{lh}(M)$, then
$\Phi$ is a presymplectic action of $G$ on $M$).
In this case we have that,
for every $\xi\in {\bf g}$, $\Lie(\tilde\xi )\Omega = 0$ or,
what is equivalent, $\inn(\tilde\xi)\Omega \in Z^1(M)$ (it is a closed 1-form).
We denote by $\tilde{\bf g}$ the set of fundamental vector fields.

Now, following the same terminology as for actions
of Lie groups on symplectic manifolds
\cite{AM-78}, \cite{LM-87}, \cite{Ok-87}, \cite{Wa-71},
we state:

\begin{definition}
$\Phi$ is said to be a {\rm strongly presymplectic} or {\rm Hamiltonian action}
of $G$ on $M$ iff, $\tilde{\bf g}\subseteq\vf_h(M)$
or, what is equivalent, for every $\xi\in {\bf g}$,
$\inn(\tilde\xi)\Omega$ is an exact form.
Otherwise, it is called a {\rm weakly presymplectic} or
{\rm locally Hamiltonian action} of $G$ on $M$.
\end{definition}

It is important to discuss when
a presymplectic action is strongly presymplectic.
The fundamental obstruction appears because the map $\Omega^\sharp$
is not an isomorphism and, as a consequence, we have
the following sequence of Lie algebras:
$$
0\longrightarrow\ker\,\Omega\longrightarrow\vf_h(M)\longrightarrow
\vf_h(M)/\ker\,\Omega\longrightarrow 0
$$
but $\vf_h(M)/\ker\,\Omega\simeq\Omega^\sharp (\vf_h(M))$,
then denoting $\vf_h(M)/\ker\,\Omega\equiv B_h^1(M)$,
we have that $B_h^1(M)\subset B^1(M)$
(where $B^1(M)$ is the set of exact differential 1-forms in $M$)
and it is a strict inclusion. In an analogous way we have the sequence
$$
0\longrightarrow\ker\,\Omega\longrightarrow\vf_{lh}(M)\longrightarrow
\vf_{lh}(M)/\ker\,\Omega\longrightarrow 0
$$
but $\vf_{lh}(M)/\ker\,\Omega\simeq\Omega^\sharp (\vf_{lh}(M))$,
then denoting $\vf_{lh}(M)/\ker\,\Omega\equiv Z_h^1(M)$,
we have that $Z_h^1(M)\subset Z^1(M)$
(where $Z^1(M)$ are the closed differential 1-forms in $M$),
and this is also a strict inclusion.
There is no problem with these exact sequences and the morphisms
relating them, but it is not possible to identify
$\vf_{lh}(M)/\vf_h(M)$ with $H^1(M)$
(the first de Rham's cohomology group of $M$),
like in the symplectic case. Nevertheless, we have
$$
\vf_{lh}(M)/\vf_h(M)\simeq
(\vf_{lh}(M)/\ker\,\Omega )/(\vf_h(M)/\ker\,\Omega )\simeq
Z_h^1(M)/B_h^1(M)
$$
and we can write
\beq
\begin{array}{ccccccccc}
0&\longrightarrow&[{\bf g},{\bf g}]&\longrightarrow&
{\bf g}\ &\longrightarrow&{\bf g}/[{\bf g},{\bf g}]&\longrightarrow&0
\\
& &\Big\downarrow \ \Xi& &\Big\downarrow \ \Xi& &\Big\downarrow \ \tilde\Xi& &
\\
0&\longrightarrow&\vf_h(M)&\longrightarrow&\vf_{lh}(M)&\longrightarrow
&Z_h^1(M)/B_h^1(M)&\longrightarrow&0
\\
& & &
\begin{picture}(24,16)(0,0)
\put(0,16){\vector(3,-2){24}}
\end{picture} &
\Big\downarrow\ \rho &
\begin{picture}(24,16)(0,0)
\put(0,0){\vector(3,2){24}}
\end{picture} & & &
\\
& & & &\vf_{lh}(M)/\ker\,\Omega& & & &
\end{array}
\label{diag}
\eeq
where $\tilde\Xi$ is a Lie algebra homomorphism
which makes the diagram commutative.
Then, the action is strongly presymplectic
(that is, the image of ${\bf g}$ by $\Xi$ is in $\vf_h(M)$)
if and only if $\tilde\Xi = 0$.
Obviously, if $H^1(M)=0$, then $Z^1(M)=B^1(M)$ and
$\vf_h(M)=\vf_{lh}(M)$, therefore $Z^1_h(M)=B^1_h(M)$.

In particular, if $(M,\Omega )$ is an exact presymplectic manifold
(that is, there exists $\Theta\in\df^1(M)$
such that $\d\Theta = \Omega$) and $\Phi$ is an exact action
(that is, $\Phi_g^*\Theta = \Theta$, for every $g\in G$)
then $\Phi$ is strongly presymplectic.

\subsection{Momentum mapping}
\protect\label{mm}

Let $G$ be a Lie group, $(M,\Omega )$ a presymplectic manifold and
$\Phi\colon G\times M \to M$ a presymplectic action of $G$ on $M$.

\begin{definition}
\ben
\item
A {\rm comomentum mapping} associated with $\Phi$ \cite{So-69}
is a Lie algebra map (if it exists)
$$
\begin{array}{ccccc}
{\cal J}^*&\colon&{\bf g}&\to&\Cinfty_h(M)
\\
& &\xi&\mapsto&f_\xi
\end{array}
$$
such that $\inn(\tilde\xi)\Omega =\d f_\xi$;
or, what is equivalent, such that  the following diagram commutes
\beq
\begin{array}{cccccccc}
& &\ \ {\bf g}&\mapping{\Xi}&\vf_{lh}(M)& & &
\\
& &{\cal J}^*\ \Big\downarrow& &
\Big\downarrow\ \rho &
\begin{picture}(24,16)(0,0)
\put(0,16){\vector(3,-2){24}}
\end{picture} & &
\\
0\longrightarrow&\Real\longrightarrow&
\Cinfty_h(M)&\mapping{\tilde\Omega^{-1}\circ\d}&\vf_{lh}(M)/\ker\,\Omega&
\longrightarrow& Z_h^1(M)/B_h^1(M)&\longrightarrow 0
\end{array}
\label{com}
\eeq
\item
A {\rm momentum mapping} associated with $\Phi$
is the dual map of a comomentum mapping; in other words, it is a map
${\cal J}\colon M\to{\bf g}^*$
such that, for every $\xi\in{\bf g}$ and $x\in M$,
$$
({\cal J}(x))(\xi ):={\cal J}^*(\xi)(x)=f_\xi(x)
$$
\een
\end{definition}

As in the symplectic case we have:

\begin{prop}
A comomentum map and the dual momentum map
associated with the presymplectic action $\Phi$ on $(M,\Omega )$ exist
if, and only if, the action is strongly presymplectic.
\end{prop}
\proof
In fact; by definition, if a comomentum mapping exists, then
$\tilde\Omega^{-1}\circ\d\circ{\cal J}^*=\rho\circ\Xi$ (see (\ref{com})),
but then
${\rm Im}\, (\rho\circ\Xi )\subset{\rm Im}\,\tilde\Omega^{-1}=
\vf_h(M)/\ker\,\Omega$,
and this implies that ${\rm Im}\,\Xi\subset\vf_h(M)$
and the action is strongly presymplectic.

Conversely, if the action is strongly presymplectic:
${\rm Im}\,\Xi\subset\vf_h(M)$, then we have that
$(\rho\circ\Xi )({\bf g})\subset \vf_h(M)/\ker\,\Omega$
and then, for all $\xi\in{\bf g}$, there exists a unique (except constants)
$f_\xi\in\Cinfty_h(M)$ such that
$\inn(\tilde\xi )\Omega =\d f_\xi$,
and this is a Lie algebra homomorphism.
\qed

Therefore, if a comomentum mapping exists,
the commutative part of the diagram (\ref{com}) reduces to
$$
\begin{array}{ccc}
\ \ {\bf g}&\mapping{\Xi}&\vf_h(M)
\\
{\cal J}^*\ \Big\downarrow& &\Big\downarrow\ \rho
\\
\Cinfty_h(M)&\mapping{\tilde\Omega^{-1}\circ\d}&\vf_{h}(M)/\ker\,\Omega
\end{array}
$$

As in the symplectic case, it is important
to point out that, if a comomentum map ${\cal J}^*$
exists for a presymplectic action,
and ${\cal F}\colon {\bf g} \to \Real$ is a linear map
(that is, ${\cal F}\in {\bf g}^*$), then
${\cal J}^*+{\cal F}$ is another comomentum map for the same action $\Phi$.
Moreover, a comomentum map is not necessarily a Lie algebra homomorphism.
Then:

\begin{definition}
The action $\Phi$ is said to be a {\rm Poissonian}
or {\rm strongly Hamiltonian action} iff:
\ben
\item
There exists a comomentum mapping for this action (and then also
a momentum one).
\item
It is a Lie algebra homomorphism.
\een
\end{definition}

As a particular case, we have that,
if $(M,\Omega )$ is an exact presymplectic manifold with $\Omega =\d\Theta$,
and the action $\Phi$ of $G$ on $M$ is exact, then:
\ben
\item
A momentum mapping exists and it is given by
${\cal J}(\xi)=-\Theta (\tilde\xi )=-\inn (\tilde\xi )\Theta$,
for every $\xi\in{\bf g}$.
\item
The action is Poissonian.
\een
In fact, the first item is immediate.
For the second one we have
$$
f_{[\xi_1,\xi_2]}=
-\inn([\tilde\xi_1,\tilde\xi_2])\Theta =
-\Lie(\tilde\xi_1)\inn(\tilde\xi_2)\Theta =
\Lie(\tilde\xi_1)f_{\xi_2}=
\{ f_{\xi_1},f_{\xi_2}\}
$$
In other cases, local comomentum mappings
can always be defined for every presymplectic action,
although without necessarily being Lie algebra homomorphisms.

In addition, we have that
if $G$ is a connected Lie group and $\Phi$ is a strongly presymplectic action
of $G$ on the presymplectic manifold $(M,\Omega )$.
Then the following statements are equivalent:
\begin{enumerate}
\item
The momentum mapping associated with this action
is $Ad^*$-equivariant, that is,
for every $g\in G$, the following diagram commutes:
\beq
\begin{array}{ccccc}
&M&\longrightarrow&{\bf g}^*&
\\
& & {\cal J} & &
\\
\Phi_g &\Big\downarrow& &\Big\downarrow& {\rm Ad}_g^*
\\
& & {\cal J} & &
\\
&M&\longrightarrow&{\bf g}^*&
\end{array}
\label{adequiv}
\eeq
\item
The action is Poissonian.
\end{enumerate}
(The proof of this statement is the same as for the symplectic case
and can be found in any of the above mentioned references).

\subsection{Level sets of the momentum mapping}
\protect\label{lsmm}

First remember that, if $\Phi$ is a strongly presymplectic action
of a Lie group $G$ on $(M,\Omega )$ and
${\cal J}$ is the momentum mapping associated to this action, then
$\mu \in {\bf g}^*$ is a {\sl weakly regular value} of ${\cal J}$ iff
\ben
\item
${\cal J}^{-1}(\mu )$ is a submanifold of $M$.
\item
$\Tan_x({\cal J}^{-1}(\mu ))=\ker\,\Tan_x{\cal J}$,
for every $x\in{\cal J}^{-1}(\mu )$.
\een
If $\Tan_x{\cal J}$ is surjective, $\mu$ is said to be a
{\sl regular value}. Of course, every regular value is weakly regular.

Taking into account that if a fundamental vector field belongs to
$\ker\,\Omega$ its Hamiltonian function can be taken to be zero,
we have that:

\begin{prop}
If $\mu$ is a weakly regular value  of ${\cal J}$ then
$\mu (\xi)=0$, for every $\xi\in{\bf g}$ such that
$\tilde\xi\in\tilde{\bf g}\cap\ker\,\Omega$.
\end{prop}

   From now on we will assume $\mu\in{\bf g}^*$ is, at least,
a weakly regular value of ${\cal J}$. So, we will denote by
$j_\mu\colon{\cal J}^{-1}(\mu )\hookrightarrow M$
the corresponding imbedding.
Then, in order to make a description of ${\cal J}^{-1}(\mu )$,
we have that the constraints defining it
are the component functions of ${\cal J}=\mu$.
In fact, observe that, if $\{\xi_i\}$ is a basis of ${\bf g}$,
$\{ f_{\xi_i}\}$ are the Hamiltonian functions
associated to the fundamental vector fields $\{\tilde\xi_i\}$
by the comomentum map and $\{ \alpha^i \}$ is the dual basis in ${\bf g}^*$,
then $\mu =\mu_i\alpha^i$, with $\mu_i$ real numbers, and we have that
\beann
{\cal J}^{-1}(\mu )&:=&\{ x\in M\ | \ {\cal J}(x)=\mu \} =
\{ x\in M\ |\ ({\cal J}(x))(\xi )=\mu (\xi )\ ,\ \forall\xi\in{\bf g} \}
\\ &=&
\{ x\in M\ |\ ({\cal J}(x))(\xi_i)=\mu_i\} =
\{ x\in M\ |\ f_{\xi_i}(x)=\mu_i\}
\eeann
that is, $j_\mu^*f_{\xi_i}-\mu_i=0$,
and then the constraints are $\zeta_i:=f_{\xi_i}-\mu_i$.
Notice that this is equivalent to saying that the expression of the
momentum mapping is
\beq
{\cal J}(x)\equiv f_{\xi_i}(x)\alpha^i
\label{expmm}
\eeq

Bearing in mind a well known result in the
theory of exterior differential systems
(see, for instance, \cite{BCG-91}),
we have that all the level sets of the momentum mapping
can also be obtained as the integral submanifolds of a Pfaff system. In fact:

\begin{prop}
Let $G$ be a Lie group and $\Phi$ a strongly presymplectic action
of $G$ on the presymplectic manifold $(M,\Omega )$.
The connected components of the level sets of the momentum mapping
${\cal J}$ associated with this
action are the connected maximal integral submanifolds of the Pfaff system
$\inn (\tilde\xi )\Omega =0$, for $\tilde\xi\in\tilde{\bf g}$.
\end{prop}

As a consequence of this proposition, we obtain that:

\begin{corol}
If $x\in{\cal J}^{-1}(\mu)$ then
$\Tan_x{\cal J}^{-1}(\mu)=\tilde{\bf g}_x^\bot$.
As a consequence, since $\ker\,\Omega_x\subset\tilde{\bf g}_x^\bot$,
then $\ker\,\Omega\subset\underline{\vf ({\cal J}^{-1}(\mu))}$
(where $\underline{\vf ({\cal J}^{-1}(\mu ))}$ denotes the set
of vector fields of $\vf (M)$ which are tangent to
${\cal J}^{-1}(\mu )$).
\end{corol}

If $\Omega =\d\Theta$ and the action is exact,
then $f_\xi=-\inn (\tilde\xi)\Theta$ and Pfaff system
$\inn (\tilde\xi )\Omega =0$
can be equivalently expressed as $\d\inn (\tilde\xi)\Theta =0$.

   From now on, we will assume the following:

\begin{assum}
The action $\Phi$ that we will consider will be
Poissonian, free and proper and $\mu$ will be a weakly regular value of the
momentum mapping associated to this action.
\end{assum}

Let $G_{\mu}$ be the isotropy group of
$\mu$ for the coadjoint action of $G$ on ${\bf g}^*$.
Then we have:

\begin{teor}
$G_\mu$ is the maximal subgroup of $G$ which lets
${\cal J}^{-1}(\mu )$ invariant and
so the quotient ${\cal J}^{-1}(\mu )/G_\mu$
is well defined and it is called the {\rm reduced phase space} or the
{\rm orbit space} of ${\cal J}^{-1}(\mu )$.
\end{teor}
\proof
For every $x\in M$ such that ${\cal J}(x)=\mu$ and $g\in G_{\mu}$, we have
$$
{\cal J}(\Phi_g(x))=({\cal J}\circ\Phi_g)(x)=
({\rm Ad}^*_g\circ{\cal J})(x)={\rm Ad}^*_g(\mu )=\mu
$$
then $\Phi_g(x)\in{\cal J}^{-1}(\mu )$,
so ${\cal J}^{-1}(\mu )$ is invariant under the action of $G_{\mu}$
and the quotient is well defined.

The maximality of $G_\mu$ is a direct consequence of the equivariance of ${\cal J}$.
\qed

If ${\bf g}_\mu$ is the Lie algebra of $G_\mu$ then,
as a consequence of this theorem,
$\tilde{\bf g}_\mu$ are vector fields tangent to ${\cal J}^{-1}(\mu )$,
and we have that
$\tilde{\bf g}_\mu =\tilde{\bf g}\cap\underline{\vf ({\cal J}^{-1}(\mu ))}$.

At this point, it is interesting to point out two different possibilities:
\bit
\item
$\tilde{\bf g}\cap\ker\,\Omega =\{ 0\}$\ :\quad
In this case all the fundamental vector fields
give constraints which are not constant functions. Then
${\rm dim}\,{\cal J}^{-1}(\mu )<m={\rm dim}\, M$.
\item
$\tilde{\bf g}\cap\ker\,\Omega\not=\{ 0\}$\ :\quad
Now, only those fundamental vector fields such that
$\tilde\xi\not \in\ker\,\Omega$ give constraints
which are not constant functions.
Then ${\rm dim}\,{\cal J}^{-1}(\mu )\leq m$.
\eit

${\cal J}^{-1}(\mu)$ inherits a presymplectic structure
$\Omega_\mu :=j_\mu^*\Omega$. We are going to characterize
$\ker\,\Omega_\mu$. First of all we have that
$\tilde{\bf g}_\mu\subset\underline{\vf ({\cal J}^{-1}(\mu ))}$
(and hence, for every $\tilde\xi\in\tilde{\bf g}_\mu$, there exists
$\tilde\xi_\mu\in\vf ({\cal J}^{-1}(\mu ))$ such that
$j_{\mu *}\tilde\xi_\mu=\tilde\xi\vert_{{\cal J}^{-1}(\mu )}$).
Consider now the {\sl orthogonal presymplectic complement} of
$\underline{\vf ({\cal J}^{-1}(\mu ))}$ in $\vf (M)$, that is, the set
\beann
(\vf ({\cal J}^{-1}(\mu )))^{\bot}&:=&
\{ Z\in\vf (M) \ \vert\ (\inn (X)\inn (Z)\Omega )(x)=0\ ,\
\forall X\in\underline{\vf ({\cal J}^{-1}(\mu ))}\ ,\
\forall x\in{\cal J}^{-1}(\mu )\}
\\ &=&
\{ Z\in\vf (M) \ \vert\ j_\mu^*\inn (Z)\Omega =0\}
\eeann
Then, let
$$
\ker\,\Omega_\mu:=\{ Y_\mu\in\vf ({\cal J}^{-1}(\mu)) \ \vert\
\inn (Y_\mu)\Omega_\mu= 0\}
$$
and denoting by $\underline{\ker\,\Omega_\mu}$ the set of vector fields of
$\vf (M)$ such that
$\underline{\ker\,\Omega_\mu}\vert_{{\cal J}^{-1}(\mu)}=
j_{\mu *}\ker\,\Omega_\mu$,
it is immediate to prove that
$$
\underline{\ker\,\Omega_\mu}=
\underline{\vf ({\cal J}^{-1}(\mu))}\cap (\vf({\cal J}^{-1}(\mu)))^{\bot}
$$

Therefore, we have the following result:

\begin{prop}
$\ker\,\Omega_{\mu_x}=\tilde{\bf g}_{\mu_x} +\ker\,\Omega_x$,
for every $x\in{\cal J}^{-1}(\mu)$.
\label{kom}
\end{prop}
\proof
For the proof see the appendix with the following identifications:
$E=\Tan_xM$, $S=\tilde{\bf g}_x$,
$N=S^{\bot}=\tilde{\bf g}_x^{\bot}=\Tan_x{\cal J}^{-1}(\mu)$,
and
$S\cap N=\tilde{\bf g}_x\cap\Tan_x{\cal J}^{-1}(\mu)=\tilde{\bf g}_{\mu_x}$.
\qed

At this point, we can state the following result which
generalizes the idea of the
{\it Marsden-Weinstein reduction theorem} \cite{MW-rsms}, \cite{We-lsm}
to presymplectic actions of Lie groups on presymplectic manifolds:

\begin{teor}
The orbit space
${\cal J}^{-1}(\mu )/G_\mu$ is a differentiable manifold.
Then, if $\sigma\colon{\cal J}^{-1}(\mu )\to{\cal J}^{-1}(\mu )/G_\mu$
denotes the canonical projection,
there is a closed 2-form $\hat\Omega\in\df^2({\cal J}^{-1}(\mu )/G_\mu)$
such that $\Omega_\mu =\sigma^*\hat\Omega$
(that is, $\Omega_\mu$ is $\sigma$-projectable), and:
\bit
\item
$\hat\Omega$ is symplectic if, and only if,
for every $x\in{\cal J}^{-1}(\mu)$,
$\tilde{\bf g}_{\mu_x} =\ker\,\Omega_{\mu_x}$
or, what is equivalent,
$\ker\,\Omega_x\cap\Tan_x{\cal J}^{-1}(\mu )\subseteq\tilde{\bf g}_{\mu_x}$.
\item
Otherwise $\hat\Omega$ is presymplectic. In particular,
for every $x\in{\cal J}^{-1}(\mu)$, if
$\ker\,\Omega_x\subset\Tan_x{\cal J}^{-1}(\mu )$
and $\tilde{\bf g}_x\cap\ker\,\Omega_x =\{ 0\}$, then
${\rm rank}\,\hat\Omega ={\rm rank}\,\Omega$.
\eit
\label{MWt}
\end{teor}
\proof
For the proof of the first part of this statement
(${\cal J}^{-1}(\mu )/G_\mu$ is a differentiable manifold)
see \cite{AM-78}, \cite{MW-rsms} or \cite{LM-87}.
The existence of $\hat\Omega$ and the two items of the second part
are a direct consequence of proposition
\ref{kom}.
\qed

\section{Symmetries of presymplectic dynamical systems and reduction}

\subsection{Review on presymplectic dynamical systems}
\protect\label{rpds}

One of the most important features in the study of
dynamical systems with symmetry is the so-called
{\it reduction theory}.
Next we want to apply the above results in order
to state the main results on this topic
concerning presymplectic dynamical systems, generalizing the ideas of the
{\it Marsden-Weinstein symplectic reduction procedure}
\cite{MW-rsms}, \cite{We-lsm}.

We start by giving the background ideas on
presymplectic dynamical systems.
For further information on this topic one may see,
for instance, \cite{GNH-pca}, \cite{CGIR-85},\cite{BK-86},
\cite{Mu-89}, \cite{Ca-90}, \cite{GP-92}
(see also \cite{Di-50} as a pioneering work).

A {\sl presymplectic locally Hamiltonian dynamical system}
is a triad $(P,\omega ,\alpha )$, where $(P,\omega )$
is a presymplectic manifold and  $\alpha\in Z^1(P)$.
If $\alpha$ is exact then $\alpha =\d{\cal H}$
for some ${\cal H}\in\Cinfty (P)$, and then the triad
$(P,\omega ,{\cal H})$ is said to be a
{\sl presymplectic Hamiltonian system}
(and this is the case we are going to consider,
without loss of generality).

Every presymplectic dynamical system has associated the following equation
$$
\inn (X_P)\omega =\d{\cal H} \quad ; \quad X_P\in\vf (P)
$$
which is compatible everywhere in $P$ if, and only if,
$\inn (Z)\d{\cal H} =0$, for every $Z\in\ker\,\omega$.
In this case $X_P\in\vf_h(P)$;
and the presymplectic dynamical system is said to be {\sl compatible}.
If the equation is not compatible, in the most interesting cases,
there is a (maximal) closed regular submanifold
$j_M \colon M \hookrightarrow P$,
for which a vector field $X_P$ tangent to $M$ exists such that
the following equation holds
\beq
[\inn(X_P)\omega -\d{\cal H}]\vert_M = 0
\label{eqres}
\eeq
(and this is an equation for $X_P$ and $M$).
$M$ is called the {\sl final constraint submanifold}
and inherits a presymplectic structure $\Omega=j_M^*\omega$.
This submanifold is obtained at the end of
a recursive algorithm which gives a sequence of submanifolds
$$
P\hookleftarrow P_1\hookleftarrow\ldots
\hookleftarrow P_{i-1}\hookleftarrow P_i\hookleftarrow\ldots
\hookleftarrow P_{f-1}\hookleftarrow P_f\equiv M
$$
The equation (\ref{eqres}) can be pulled-back to $M$ obtaining
\beq
j_M^*[\inn(X_P)\omega -\d{\cal H}]=\inn(X)\Omega-\d{\rm H}=0
\label{eqnat}
\eeq
with ${\rm H}=j_M^*{\cal H}$,
$X_P\in\vf (P)$ tangent to $M$, $X\in\vf (M)$ and $j_{M_*}X=X_P\vert_M$.
Notice that this is a compatible system because
if $X_P\in\vf (P)$ tangent to $M$ is a vector field solution of
(\ref{eqres}), this implies that the equation (\ref{eqnat})
holds for $X$.

For the final submanifold $M$ the vector field $X_P$
satisfying (\ref{eqres}) is not unique, in general.
Then the difference between two solutions is called a {\sl gauge vector field};
and the points of $M$ reached from another fixed one $x\in M$
by means of an integral curve of a gauge vector field
(passing through $x$) are the so-called {\sl gauge equivalent points} or {\sl states}.
Under certain regularity conditions, it is proved that
the set of gauge vector fields is just $\underline{\ker\,\Omega}$
(the necessary and sufficient condition for this is the following \cite{Go-79}, \cite{BK-86}:
the constraint functions locally defining $M$ in $P$ can be classified into
{\sl first} and {\sl second class}; then there is a basis of the set of first class constraints whose differentials do not vanish along $M$).

In order to remove the redundancy of solutions,
it is assumed that gauge equivalent states represent the same
physical state. Geometrically this means that
we must go from $M$ to the quotient
of $M$ by the foliation generated by the involutive distribution $\underline{\ker\,\Omega}$.
It is assumed that the quotient space $\bar M\equiv M/\ker\,\Omega$
is a differentiable manifold,
the projection $\pi_M\colon M\to\bar M$ is a submersion
and $\bar M$ is endowed with a symplectic structure $\bar\omega$
such that $\pi_M^*\bar\omega=\Omega$.
$(\bar M,\bar\omega)$ is called the {\sl manifold of real physical states},
and the equations (\ref{eqres}) and (\ref{eqnat})
project in a natural way to $\bar M$:
\beq
\inn(\bar X_P)\bar\omega-\d\bar{\rm H}=0
\label{eqred}
\eeq
where $\pi_M^*\bar{\rm H}={\rm H}$ and
for every $X_P\in\underline{\vf (M)}$ which is solution of (\ref{eqres}),
there is a $\pi_M$-projectable vector field $X\in\vf (M)$,
with $j_{M_*}X=X_P\vert_M$, such that
$\pi_{M_*}X=\bar X_P$ is the unique solution of (\ref{eqred}).
Note that the existence of $\bar{\rm H}$ is assured
because $\inn(\ker\,\Omega){\rm H}=0$,
since the dynamical equation (\ref{eqnat}) on $M$ is compatible.
This is the so-called {\sl gauge reduction procedure}.

On the other hand, the following {\sl structure theorem} for
presymplectic dynamical systems plays a crucial role in some
of the developments of this work:

\begin{teor}
Let $(M,\Omega,{\rm H})$ be a compatible presymplectic dynamical system. Then:
\begin{enumerate}
\item
There exists a symplectic manifold $({\bf M},{\bf \Omega})$
such that $j_0\colon M \hookrightarrow {\bf M}$ is a coisotropic imbedding,
and $j_0^*{\bf  \Omega}=\Omega$.
\item
There exists a family ${\cal D}_{lh}({\bf M},M)$
of symplectic locally Hamiltonian vector fields in ${\bf M}$ tangent to $M$,
which gives all the dynamical solutions of the equation
$$
j_0^*[\inn(X_\beta){\bf \Omega}-\d{\rm H}]=\inn(X)\Omega-\d{\rm H}=0
\quad ; \quad
X_\beta\in{\cal D}_{lh}({\bf M},M)
$$
\item
The symplectic manifold $({\bf M},{\bf \Omega})$
and the family ${\cal D}_{lh}({\bf M},M)$
are unique up to symplectomorphic neighbourhood equivalences
between symplectic manifolds containing $(M,\Omega)$
as a coisotropic submanifold; and all these symplectomorphisms
reduce to the identity on $M$.
(That is, if $j_i\colon M\hookrightarrow ({\bf M}_i,{\bf \Omega}_i)$, $i=1,2$,
are two coisotropic imbeddings, then there exist two
tubular neighbourhoods $U_i$ of $j_i(M)$ in ${\bf M}_i$ and a symplectomorphism
$\psi\colon ({\bf M}_1,{\bf \Omega}_1)\to ({\bf M}_2,{\bf \Omega}_2)$
such that $\psi\circ j_1=j_2$ and
$\psi_*({\cal D}_{lh}({\bf M}_1,M)={\cal D}_{lh}({\bf M}_2,M)$).
\end{enumerate}
The pair $({\bf M},{\bf \Omega})$ is called an
{\sl ambient symplectic manifold} for $(M,\Omega)$ and
$({\bf M},{\bf \Omega},{\bf H})$ is called an
{\sl ambient symplectic dynamical system} for the
presymplectic system $(M,\Omega,{\rm H})$.
\end{teor}
({\sl Outline of the proof}) \quad
The first part of this statement (together with the symplectomorphic
equivalence of coisotropic imbeddings) is the well-known
{\sl coisotropic imbedding theorem} \cite{Go-82} \cite{Ma-83}.
The symplectic manifold $({\bf M},{\bf \Omega})$
is constructed as a tubular neighbourhood of the zero section
of the dual characteristic bundle $K^*\equiv (\ker\,\Omega)^*$,
which is identified with $M$.
The strategy consists in considering the vector bundle
$\pi_K\colon K\to M$; then, as $K$ is a subbundle of $\Tan M$,
using a metric in $M$, we can split $\Tan M=G\oplus K$,
and then $\Tan_MK^*=\Tan M\oplus K^*=G\oplus K\oplus K^*$.
Denoting $\sigma\colon \Tan_MK\to K\oplus K^*$, we set
${\bf  \Omega}=\pi_K^*\Omega+\sigma^*\Omega_K$;
where $\Omega_K$ is the symplectic form canonically defined in $K\oplus K^*$.
Then ${\bf \Omega}$ can be extended to a tubular neighbourhood of $M$ in
$K^*$ using {\sl Weinstein's extension theorem} \cite{We-lsm}.

In relation to the second part,
the family ${\cal D}_{lh}({\bf M},M)$ is made of the vector fields
$X_{\beta}={\bf \Omega}^{-1}(\d{\bf H}+\beta)$,
where ${\bf H}\in\Cinfty ({\bf M})$ is an extension of ${\rm H}$ to ${\bf M}$
(that is, $j_0^*{\bf H}={\rm H}$)
and $\beta\in Z^1({\bf M})$ is a closed first class constraint one form,
(that is, $j_0^*\beta=0$ and
$j_0^*\inn (Z)\beta=0$, $\forall Z\in\vf (M)^\bot$).
(See \cite{CGIR-85} for the details of this part of the proof).

Finally, the local uniqueness of the coisotropic imbedding is a
straightforward consequence of the local uniqueness part of
Weinstein's extension theorem.
\qed

   From now on we will consider presymplectic dynamical systems
$(P,\omega,{\cal H})$ with final constraint submanifold $(M,\Omega,{\rm H})$
which hold all these features.

\subsection{Reduction of compatible presymplectic dynamical systems
with symmetry}
\protect\label{spds}

Consider now a compatible presymplectic dynamical system
$(M,\Omega,{\rm H})$; that is, such that the dynamical equation
\beq
\inn (X)\Omega -\d{\rm H}=0
\label{eqfundam}
\eeq
has solution $X\in\vf (M)$ everywhere in $M$.
We are then able to introduce the concept of
{\sl group of symmetries} for a compatible presymplectic dynamical system
(and its reduction) as follows
(the case of non-compatible systems will be studied afterwards):

\begin{definition}
Let $G$ be a Lie group, $(M,\Omega,{\rm H})$ a compatible presymplectic
dynamical system and $\Phi\colon G\times M\to M$ an action of $G$ on $M$.
$G$ is said to be a {\rm symmetry group} of this system iff
\ben
\item
$\Phi$ is a presymplectic action on $(M,\Omega )$
\item
$\Phi_g^*{\rm H}={\rm H}$; for every $g\in G$ or,
what is equivalent,
$\Lie(\tilde\xi ){\rm H}=0$, for every $\xi\in{\bf g}$.
\een
The diffeomorphism
$\Phi_g$, for every $g\in G$, is called a {\rm symmetry} of the system.
The fundamental vector fields $\tilde\xi\in\tilde{\bf g}$ are the so-called
{\rm infinitesimal generators of symmetries}.
\label{simg}
\end{definition}

Obviously this definition is the same as
the usual one for symmetries of symplectic dynamical systems.
At this point, we first prove that
gauge symmetries are symmetries of the presymplectic dynamical system.

\begin{prop}
Let $(M,\Omega ,{\rm H})$ be a compatible presymplectic dynamical system.
Then the vector fields of $\ker\,\Omega$ are infinitesimal generators
of symmetries of this system.
\label{Ginker}
\end{prop}
\proof
First, for every $Z\in\ker\,\Omega$,
by definition $\inn (Z)\Omega =0$ and hence $\Lie (Z)\Omega =0$.
On the other hand, since the equation (\ref{eqfundam}) is compatible,
it implies that
$$
\inn (Z)\d{\rm H}=\Lie (Z){\rm H}=0
$$
\qed

According to the terminology of the above section, we say that
the vector fields of $\ker\,\Omega$ are infinitesimal generators of symmetries
of the system. Then, if we want to remove the symmetries, following a
reduction procedure in order to get a symplectic dynamical system,
we must suppose that the vector fields in $\ker\,\Omega$
are contained in the distribution generated by $\tilde{\bf g}$.
So, from now on we will assume that:

\begin{assum}
Let $\tilde{\bf g}$ be the vector space made of the
fundamental vector fields of the action $\Phi$ of the symmetry group $G$
on $M$. Then
$\ker\,\Omega\subset\Cinfty (M)\otimes\tilde{\bf g}$.
\label{asum2}
\end{assum}

{\bf Comments}:
\bit
\item
This assumption means that, if
$\{\moment{\xi}{1}{h}\}\subset{\bf g}$ is a basis of ${\bf g}$,
and $Z\in\ker\,\Omega$; then there exist
 $\{\coor{f^i}{1}{h}\}\subset\Cinfty (M)$
such that $Z=f^i\tilde\xi_i$.
\item
Observe that $\Cinfty (M)\otimes\tilde{\bf g}$ is the submodule of
$\vf (M)$ made of the vector fields tangent to the orbits of the action of $G$
(or, what is equivalent, if the distribution defined by $\tilde{\bf g}$ has
constant dimension, they are the sections of this distribution).
Therefore, the assumption means that the elements of $\ker\,\Omega$
are tangent to these orbits. Hence, the leaves of the foliation
induced by $\ker\,\Omega$ are contained in those orbits.
\item
Notice that the elements of $\ker\,\Omega$ are infinitesimal generators of symmetries
but , in general, those of $\Cinfty (M)\otimes\tilde{\bf g}$ are not.

In the usual physical terminology, the vector fields of $\ker\,\Omega$
are called infinitesimal generators of {\sl gauge symmetries}.
On the contrary, the vector fields of $\tilde{\bf g}$ which do not
belong to $\ker\,\Omega$ are the so-called
infinitesimal generators of {\sl non-gauge} or {\sl rigid symmetries}.
\eit

Now, suppose that the action of the symmetry group $G$ on the
compatible presymplectic dynamical system $(M,\Omega,{\rm H})$ is Poissonian.
Let ${\cal J}$ be the momentum mapping associated with this action,
and $\mu\in{\bf g}^*$ a weakly regular value of ${\cal J}$.
Then the submanifold ${\cal J}^{-1}(\mu )$,
the form $\Omega_\mu =j_\mu^*\Omega$
and the function ${\rm H}_\mu =j_\mu^*{\rm H}$ make
a presymplectic Hamiltonian system
$({\cal J}^{-1}(\mu ),\Omega_\mu ,{\rm H}_\mu)$.
Then we have:

\begin{prop}
If $X\in\vf_h(M)$ is a vector field solution of the
equation (\ref{eqfundam}), then:
\ben
\item
$X$ is tangent to ${\cal J}^{-1}(\mu )$.
\item
The dynamical equation
\beq
\inn(X_\mu)\Omega_\mu -\d{\rm H}_\mu =0
\label{ecotra}
\eeq
is compatible and its solutions are $X_\mu +\ker\,\Omega_\mu$,
where $X_\mu\in\vf ({\cal J}^{-1}(\mu ))$ is a vector field
such that $j_{\mu *}X_\mu =X\vert_{{\cal J}^{-1}(\mu )}$.
\een
\label{shred}
\end{prop}
\proof
First we prove that if $X\in\vf (M)$ is a solution of
the equation (\ref{eqnat}), then
$X\in\underline{\vf ({\cal J}^{-1}(\mu ))}$.
In fact, for every constraint $\zeta$,
with $\d\zeta =\inn (\tilde\xi)\Omega$, $\tilde\xi\in\tilde{\bf g}$,
defining ${\cal J}^{-1}(\mu )$,
$$
j_\mu^*X(\zeta )=
j_\mu^*(\inn (X)\d\zeta )=
j_\mu^*(\inn (X)\inn (\tilde\xi)\Omega )=
-j_\mu^*(\inn (\tilde\xi)\inn (X)\Omega )=
-j_\mu^*(\inn (\tilde\xi)\d{\rm H})=0
$$
Therefore
$$
\inn(X_\mu)\Omega_\mu -\d{\rm H}_\mu =
j_\mu^*(\inn(X)\Omega -\d{\rm H})=0
$$
and the second result follows.
\qed

In addition we have that:

\begin{lem}
$\ker\,\Omega\subset\underline{\vf ({\cal J}^{-1}(\mu ))}$.
(That is, $\ker\,\Omega$ lets ${\cal J}^{-1}(\mu )$ invariant).
\end{lem}
\proof
In fact, take the constraint functions $\{\zeta\}$ defining
${\cal J}^{-1}(\mu )$ such that
$\d\zeta =\inn (\tilde\xi )\Omega$, for some $\tilde\xi\in\tilde{\bf g}$.
Then, if $Z\in\ker\,\Omega$, we have that
$$
\Lie(Z)\zeta =\inn(Z)\d\zeta =\inn(Z)\inn(\tilde\xi )\Omega =
-\inn(\tilde\xi )\inn(Z)\Omega =0
$$
therefore $Z\in\underline{\vf ({\cal J}^{-1}(\mu ))}$.
\qed

\begin{lem}
$\tilde{\bf g}_{\mu_x}=\ker\,\Omega_{\mu_x}$;
for every $x\in{\cal J}^{-1}(\mu)$.
\end{lem}
\proof
By proposition \ref{kom} we have that
$\ker\,\Omega_{\mu_x}=\tilde{\bf g}_{\mu_x} +\ker\,\Omega_x$,
and the assumption \ref{asum2} gives us that
$\ker\,\Omega_x\subset\tilde{\bf g}_x$.
On the other hand, $\tilde{\bf g}_{\mu_x}$ is the maximal subspace of
$\tilde{\bf g}_x$ being tangent to  ${\cal J}^{-1}(\mu)$;
and $\ker\,\Omega_x$ is made of vectors which are tangent to ${\cal J}^{-1}(\mu)$
(as a consequence of the above lemma).
Therefore $\ker\,\Omega_x\subset \tilde{\bf g}_{\mu_x}$,
and the result follows.
\qed

Now, the last step is to obtain the orbit space
$({\cal J}^{-1}(\mu )/G_\mu ,\hat\Omega )$.

\begin{teor}
Consider the presymplectic Hamiltonian system
$({\cal J}^{-1}(\mu ),\Omega_\mu ,{\rm H}_\mu)$,
the quotient manifold ${\cal J}^{-1}(\mu )/\ker\,\Omega_\mu$,
and the canonical projection
$\pi_\mu\colon {\cal J}^{-1}(\mu )\to{\cal J}^{-1}(\mu )/\ker\,\Omega_\mu$.
Then the function ${\rm H}_\mu$ and the vector field
$X_\mu\in\vf ({\cal J}^{-1}(\mu ))$ of the proposition \ref{shred}
are $\pi_\mu$-projectable.
Hence $({\cal J}^{-1}(\mu )/\ker\,\Omega_\mu ,\hat\Omega ,\hat{\rm H})$,
is a symplectic Hamiltonian system and
\beq
\inn(\hat X)\hat\Omega -\d\hat{\rm H}=0
\label{ecuac}
\eeq
where $\pi_\mu^*\hat{\rm H}={\rm H}_\mu$ and $\pi_{\mu*}X_\mu =\hat X$.
\end{teor}
\proof
According to the last lemma,
we have that $\tilde{\bf g}_{\mu_x}=\ker\,\Omega_{\mu_x}$,
for every $x\in{\cal J}^{-1}(\mu)$, and so
${\cal J}^{-1}(\mu)/G_\mu={\cal J}^{-1}(\mu)/\ker\,\Omega_\mu$.
Then, taking into account the first item of theorem \ref{MWt} we have that
$({\cal J}^{-1}(\mu )/\ker\,\Omega_\mu,\hat\Omega)$ is a symplectic manifold.

Now, in order to see that ${\rm H}_\mu$ is $\pi_\mu$-projectable it suffices to
prove that $\Lie(\tilde\xi_\mu){\rm H}_\mu =0$,
for every $\tilde\xi_\mu\in\tilde{\bf g}_\mu\subset\tilde{\bf g}$. But this holds
since ${\rm H}$ is $G$-invariant and then ${\rm H}_\mu$ is $G_\mu$-invariant.

On the other hand, for every $\tilde\xi_\mu\in\tilde{\bf g}_\mu$, we have that
$$
\inn([\tilde\xi_\mu,X_\mu])\Omega_\mu =
\Lie(\tilde\xi_\mu)\inn(X_\mu)\Omega_\mu -
\inn(X_\mu)\Lie(\tilde\xi_\mu)\Omega_\mu =
\Lie(\tilde\xi_\mu)\d{\rm H}_\mu =0
$$
since $\Omega_\mu$ and ${\rm H}_\mu$ are $G_\mu$-invariant,
and then $[\tilde\xi_\mu,X_\mu]\in\ker\,\Omega_\mu$.
But, as all the elements of $\ker\,\Omega_\mu$ are of the form
$Z_\mu=f^i\xi_{\mu i}$, then we also have that
$[Z_\mu,X_\mu]\in\ker\,\Omega_\mu$, for every $Z_\mu\in\ker\,\Omega_\mu$;
and therefore, $X_\mu$ is $\pi_\mu$-projectable.

Finally, the equation (\ref{ecuac}) follows immediately from (\ref{ecotra}).
\qed

We can summarize the procedure in the following diagram
$$
(M,\Omega,{\rm H})\
\begin{picture}(24,5)(0,0)
\put(24,0){\vector(-1,0){24}}
\put(10,4){\mbox{$j_\mu$}}
\end{picture}
\ ({\cal J}^{-1}(\mu),\Omega_\mu,{\rm H}_\mu)\
\begin{picture}(24,5)(0,0)
\put(0,0){\vector(1,0){24}}
\put(10,4){\mbox{$\pi_\mu$}}
\end{picture}
\ ({\cal J}^{-1}(\mu)/\ker\,\Omega_\mu,\hat\Omega,\hat{\rm H})
$$
The final result is a reduced symplectic dynamical system
$({\cal J}^{-1}(\mu)/\ker\,\Omega_\mu,\hat\Omega,\hat{\rm H})$.
Observe that, making only one quotient, we have removed the symmetries of the action of $G$ and the non-uniqueness arising from the existence of $\ker\,\Omega$;
then obtaining a symplectic dynamical system where $G$ acts by the identity.

   From now on we will refer to this
reduction scheme as the {\sl complete presymplectic reduction} procedure.

\subsection{Other reduction procedures:
Gauge reduction {\it plus} symplectic reduction}
\protect\label{comp1}

We finish this study by comparing the reduction method
of presymplectic dynamical systems with symmetry
here presented with the other {\it step-by-step}
reduction procedures.
In particular, the reduction procedure that has been developed
in the last section, removes both rigid and gauge symmetries.
Now, in this section, we make these procedures successively,
proving that we obtain the same result as above.

Consider a compatible presymplectic dynamical system $(M,\Omega,{\rm H})$.
Let $G$ be a group of symmetries of the presymplectic dynamical system,
${\bf g}$ its Lie algebra and $\tilde{\bf g}$ the
corresponding algebra of fundamental vector fields.
We suppose that the assumption \ref{asum2} holds.

First, we apply the gauge reduction procedure obtaining
the reduced phase space $(\bar M,\bar\omega,\bar{\rm H})$
(which is a symplectic dynamical system),
with $\pi_M\colon M\to \bar M$.
So the gauge symmetries have been removed.
Next we must study under what conditions the action of $G$
goes down to $\bar M$ and, therefore, the corresponding
non-gauge symmetry can be removed by means of
the standard symplectic reduction procedure of Marsden-Weinstein.

\begin{prop}
If $\Phi\colon G\times M\to M$ is a (strongly) presymplectic action,
then there exists a ``reduced action'' $\bar\Phi\colon G\times \bar M\to \bar M$
such that $\bar\Phi_g(\pi_M(p)):=\pi_M(\Phi_g(p))$;
for every $g\in G$ and $p\in M$.
\end{prop}
\proof
In order to see that this reduced action $\bar\Phi$ is well defined,
we must prove that, given $p_1,p_2\in M$ belonging to the same leaf
of the foliation defined by $\ker\,\Omega$, then $\Phi_g(p_1)=\Phi_g(p_2)$,
for every $g\in G$.
If $p_1,p_2$ are in the same leaf of this foliation, then they are
connected by a (piecewise) regular curve, which is made of
(pieces of) integral curves of vector fields belonging to $\ker\,\Omega$.
But, if $\gamma (t)$ is an integral curve of some $Z\in\ker\,\Omega$, then
$(\Phi_g\circ\gamma)(t)$ is an integral curve of $\Phi_{g_*}Z$,
which is also a vector field in $\ker\,\Omega$ since,
as $\Phi$ is a presymplectic action, we have
$$
\inn (\Phi_{g_*}Z)\Omega=\inn(\Phi_{g_*}Z)(\Phi_g^{-1})^*\Omega=
(\Phi_g^{-1})^*(\inn(Z)\Omega)=0
$$
Therefore the action $\bar\Phi$ can be defined as is set in the statement.
\qed

Now the problem is that, although the action $\Phi$ can be assumed to be free,
the reduced action $\bar\Phi$ is not so in general;
because the assumption \ref{asum2} implies that the leaves of the foliation
induced by $\ker\,\Omega$ are in the orbits of $G$ and,
then, the quotient by $\ker\,\Omega$ leads to a non free action, in general.
Hence we set the following hypothesis
(which is implicitly assumed in the physical literature):

\begin{assum}
Let $\tilde{\bf g}$ be the vector space of the
fundamental vector fields of the action $\Phi$ of the symmetry group $G$
on $M$. Then there is a subalgebra ${\bf G}\subset{\bf g}$,
such that the corresponding $\tilde{\bf G}\subset\tilde{\bf g}$ verifies that
$\ker\,\Omega\subset\Cinfty (M)\otimes\tilde{\bf G}$.
\label{asum3}
\end{assum}

Now, suppose that ${\cal G}\subset G$ (the subgroup having ${\bf G}$ as Lie algebra)
is closed; and let $\bar G:=G/{\cal G}$ be the quotient group, which acts on
$\bar M$ by the reduced action $\bar\Phi$.
Denote by ${\bf g}_{\bar M}$ the Lie algebra of $\bar G$, and by
$\tilde{\bf g}_{\bar M}$ the corresponding set of fundamental vector fields.
Then we have the projections
$$
\begin{array}{ccccc}
{\bf g}&\longrightarrow&{\bf g}_{\bar M}&\longrightarrow&{\bf 0}
\\
\xi&\mapsto&\xi^{\bar M}& &
\end{array}
$$
and the duals
\beq
\begin{array}{ccccc}
{\bf 0}&\longrightarrow&{\bf g}^*_{\bar M}&\longrightarrow&{\bf g}^*
\\
& &\bar\mu&\mapsto&\mu
\end{array}
\label{mumu}
\eeq

We have:

\begin{lem}
The set of fundamental vector fields
$\tilde{\bf g}$ is $\pi_M$-projectable.
\end{lem}
\proof
In fact, for every $\tilde\xi\in\tilde{\bf g}$ and
for every $Z\in\ker\,\Omega$, we have that
$$
\inn([\tilde\xi,Z ])\Omega =
\Lie(\tilde\xi)\inn(Z)\Omega -\inn(Z)\Lie(\tilde\xi)\Omega =0
$$
therefore $[\tilde\xi,Z]\in\ker\,\Omega$ and the result follows.
\qed

And therefore:

\begin{prop}
With the above assumptions, if the action $\Phi$
of $G$ on $M$ is strongly presymplectic and free, then
the reduced action $\bar\Phi$ of the quotient group $\bar G$ on $\bar M$
is strongly symplectic and free.
\label{propfree}
\end{prop}
\proof
First we prove that $\bar\Phi$ is free. In fact;
if $\bar g\in\bar G$ and $p\in M$, then $\bar\Phi_{\bar g}(\pi_M(p))=\pi_M(p)$,
by definition. But, if $\bar\Phi_g(\pi_M(p))=\pi_M(p)$, for some $g\in G$,
then $\pi_M(\Phi_g(p)=\pi_M(p)$, and hence $g\in{\cal G}$.
Therefore the isotropy group of a point of the action of $G$ on $M$ is ${\cal G}$,
thus the action of $\bar G$ on $\bar M$ is free.

Second we prove that $\bar\Phi$ is strongly symplectic. In fact;
for every $\tilde\xi\in\tilde{\bf g}$
with $\inn(\tilde\xi)\Omega =\d f_\xi$, we have that
$f_\xi$ is $\pi_M$-projectable. In fact
$$
\Lie(Z)f_\xi =\inn(Z)\d f_\xi =\inn(Z)\inn(\tilde\xi)\Omega =
-\inn(\tilde\xi)\inn(Z)\Omega =0
$$
for every $Z\in\ker\,\Omega$. Then the function
$\bar f_\xi\in\Cinfty (\bar M)$ such that $\pi_M^*\bar f_\xi=f_\xi$
is Hamiltonian for the fundamental vector field
$\tilde\xi^{\bar M}\in\tilde{\bf g}_{\bar M}$
such that $\pi_{M_*}\tilde\xi =\tilde\xi^{\bar M}$, since
$$
\pi_M^*\d\bar f_\xi =\d\pi_M^*\bar f_\xi =\d f_\xi =
\inn(\tilde\xi)\Omega =\inn(\tilde\xi)\pi_M^*\bar\omega =
\pi_M^*\inn(\tilde\xi^{\bar M})\bar\omega
$$
and hence $\d\bar f_\xi =\inn(\tilde\xi^{\bar M})\bar\omega$,
since $\pi_M$ is a submersion.
\qed

Now, we can define the {\sl reduced comomentum mapping}
associated with $\bar\Phi$ as a Lie algebra linear map
$$
\begin{array}{ccccc}
\bar{\cal J}^*&\colon&{\bf g}_{\bar M}&\to&\Cinfty (\bar M)
\\
& &\xi^{\bar M}&\mapsto&\bar f_\xi
\end{array}
$$
such that $\inn(\tilde\xi^{\bar M})\bar\omega =\d\bar f_\xi$.
The {\sl reduced momentum mapping} associated with $\bar\Phi$
is its dual map, $\bar{\cal J}\colon \bar M\to{\bf g}^*_{\bar M}$;
that is, for every $\xi^{\bar M}\in{\bf g}_{\bar M}$ and $x\in M$,
$$
(\bar{\cal J}(\pi_M(x)))(\xi^{\bar M}):=
\bar{\cal J}^*(\xi^{\bar M})(\pi_M(x))=
\bar f_\xi(\pi_M(x))
$$
Finally, it can be proved that if $\Phi$ is
a Poissonian and proper action, then so is $\bar\Phi$.

At this point, the standard symplectic reduction program is applied:
\bit
\item
First we construct the level sets
$(\bar{\cal J}^{-1}(\bar\mu),\omega_{\bar\mu},{\rm H}_{\bar\mu})$, for
weakly regular values $\bar\mu\in{\bf g}_{\bar M}^*$.
Each one of them is locally defined in $(\bar M,\bar\omega)$
by the constraints $\{\bar f_\xi\}$.
\item
Second we take the isotropy group
$\bar G_{\bar\mu}$ and its Lie algebra $({\bf g}_{\bar M})_{\bar\mu}$,
for which we have that
$(\tilde{\bf g}_{\bar M})_{\bar\mu}=\ker\,\omega_{\bar\mu}$.
Then we make the quotient $\bar{\cal J}^{-1}(\bar\mu )/\bar G_{\bar\mu}$.
\eit
After that, the reduced symplectic Hamiltonian system
$(\bar{\cal J}^{-1}(\bar\mu)/\ker\,\omega_{\bar\mu},\hat\omega,\hat{\rm h})$
is free of gauge and rigid symmetries
and the following theorem proves that it coincides with
$({\cal J}^{-1}(\mu)/\ker\,\Omega_\mu,\hat\Omega,\hat{\rm H})$,
which is the one obtained after the complete presymplectic reduction procedure
(for $\mu$ and $\bar\mu$ related as shown in (\ref{mumu})).

\begin{teor}
There exists a diffeomorphism
$$
\rho\colon
({\cal J}^{-1}(\mu)/\ker\,\Omega_\mu,\hat\Omega,\hat{\rm H})
\longrightarrow
(\bar{\cal J}^{-1}(\bar\mu)/\ker\,\omega_{\bar\mu},\hat\omega,\hat{\rm h})
$$
such that $\hat\Omega=\rho^*\hat\omega$
and $\hat{\rm H}=\rho^*\hat{\rm h}$.
\end{teor}
\proof
First of all, we can construct a map
$\tau\colon
({\cal J}^{-1}(\mu),\Omega_\mu,{\rm H}_\mu)\to
(\bar{\cal J}^{-1}(\bar\mu),\omega_{\bar\mu},{\rm H}_{\bar\mu})$
verifying the relation
$j_{\bar\mu}\circ\tau =\pi_M\circ j_\mu$
(see the diagram below).
Observe that, if $p\in{\cal J}^{-1}(\mu)$
then $\pi_M(p)\in\bar{\cal J}^{-1}(\bar\mu)$,
by the proposition \ref{propfree} and the relation between $\mu$ and $\bar\mu$
(see (\ref{mumu})); hence $\tau$ is well defined and it is a
surjective submersion (since so is $\pi_M$).
Moreover, we have that
$$
\Omega_\mu=j_\mu^*\pi_M^*\bar\omega=
\tau^*j_{\bar\mu}^*\bar\omega=
\tau^*\omega_{\bar\mu}
$$
and in the same way we obtain that ${\rm H}_\mu=\tau^*{\rm H}_{\bar\mu}$.

Now there is an unique map
$\rho\colon
({\cal J}^{-1}(\mu)/\ker\,\Omega_\mu,\hat\Omega,\hat{\rm H})
\longrightarrow
(\bar{\cal J}^{-1}(\bar\mu)/\ker\,\omega_{\bar\mu},\hat\omega,\hat{\rm h})$
such that $\rho\circ\pi_\mu =\tau\circ\pi_{\bar\mu}$.
So we have the diagram
$$
\matrix{complete \cr presymplectic \cr reduction}
\left\{
\begin{array}{ccc}
(M,\Omega,{\rm H}) &
\begin{picture}(24,5)(0,0)
\put(0,0){\vector(1,0){24}}
\put(10,4){\mbox{$\pi_M$}}
\end{picture}
& (\bar M,\bar\omega,\bar{\rm H})
\\
j_\mu\ \Big\uparrow & & \Big\uparrow\ j_{\bar\mu}
\\
({\cal J}^{-1}(\mu),\Omega_\mu,{\rm H}_\mu) &
\begin{picture}(24,5)(0,0)
\put(0,0){\vector(1,0){24}}
\put(10,4){\mbox{$\tau$}}
\end{picture}
& (\bar{\cal J}^{-1}(\bar\mu),\omega_{\bar\mu},{\rm H}_{\bar\mu})
\\
\pi_\mu\ \Big\downarrow & & \Big\downarrow\ \pi_{\bar\mu}
\\
({\cal J}^{-1}(\mu)/\ker\,\Omega_\mu,\hat\Omega,\hat{\rm H}) &
\begin{picture}(24,5)(0,0)
\put(0,0){\vector(1,0){24}}
\put(10,4){\mbox{$\rho$}}
\end{picture}
& (\bar{\cal J}^{-1}(\bar\mu)/\ker\,\omega_{\bar\mu},\hat\omega,\hat{\rm h})
\end{array}
\right\}
\matrix{standard \cr symplectic \cr reduction}
$$
The map $\rho$ has the following properties:
\ben
\item
$\rho$ is well defined:

Let $p_1,p_2\in{\cal J}^{-1}(\mu)$ such that $\pi_\mu(p_1)=\pi_\mu(p_2)$;
we have to prove that $\pi_{\bar\mu}\tau (p_1)=\pi_{\bar\mu}\tau (p_2)$.
Since $\pi_\mu(p_1)=\pi_\mu(p_2)$, it implies that $p_1,p_2$
can be joined by a polygonal made of integral curves of
vector fields of $\ker\,\Omega_\mu$
(but we will take a single curve, since it suffices to repeat the
reasoning a finite number of times).
For every point $p$ of the curve, if $Z_p\in\Tan_p{\cal J}^{-1}\mu$
is tangent to this curve at $p$, as $\tau$ is a surjective submersion and
$\tau^*\omega_{\bar\mu}=\Omega_\mu$, we have that
$$
\inn (Z_p)(\Omega_\mu)_p=
\inn (Z_p)(\tau^*\omega_{\bar\mu})_p=
\tau_p^*\inn (\Tan_p\tau(Z_p))(\omega_{\bar\mu})_{\pi_{\tau(p)}}=0
$$
As a consequence
$\Tan_p\tau\vert_{\ker\, (\Omega_\mu)_p}\colon
\ker\, (\Omega_\mu)_p\to\ker\, (\omega_{\bar\mu})_{\tau(p)}$
is surjective and
$\ker\,\Tan_p\tau\subset\ker\,(\Omega_\mu)_p$.
Hence, the curve joining $p_1$ and $p_2$ can be covered
by a finite number of open sets satisfying this property and,
then, $\tau(p_1)$ and $\tau(p_2)$ are connected by a polygonal
of integral curves of vector fields of $\ker\,\omega_{\bar\mu}$;
that is, $\pi_{\bar\mu}\tau (p_1)=\pi_{\bar\mu}\tau (p_2)$.
\item
$\rho$ is bijective:

$\tau$ maps the leaves of the foliation defined by
$\ker\,\Omega_\mu$ into the leaves of the foliation defined by
$\ker\,\omega_{\bar\mu}$; therefore $\rho$ is injective.
To see that $\rho$ is surjective is trivial.
\item
$\rho$ is a diffeomorphism:

$\pi_\mu\colon {\cal J}^{-1}(\mu)\to
{\cal J}^{-1}(\mu)/\ker\,\Omega_\mu$
is a surjective submersion, then there are differentiable local sections
$s_\mu\colon {\cal J}^{-1}(\mu)/\ker\,\Omega_\mu
\to {\cal J}^{-1}(\mu)$
(that is, such that $\pi_\mu\circ s_\mu={\rm Id}$);
hence, locally, we have that $\rho=\pi_{\bar\mu}\circ\tau\circ s_\mu$
and then $\rho$ is differentiable since so are
$\pi_{\bar\mu},\tau$ and $s_\mu$.
(The choice of the local sections $s_\mu$
is called {\sl gauge fixing} in the physical literature).

Now we must prove that $\rho^{-1}$ is differentiable.
Taking into account the theorem of the inverse function,
it is sufficient to prove that the tangent map $\Tan_{\pi_\mu(p)}\rho$
is an isomorphism for every $p\in{\cal J}^{-1}(\mu)$.
Then, let $u\in\Tan_{\pi_\mu(p)}({\cal J}^{-1}(\mu)/\ker\,\Omega_\mu)$
such that $\Tan_{\pi_\mu(p)}\rho(u)=0$ and
$v\in (\Tan_p\pi_\mu)^{-1}(u)$. By the commutativity
of the diagram above we have that
$\Tan_p(\pi_{\bar\mu}\circ\tau)(v)=0$; hence
$\Tan_p\tau(v)\in\ker\,\Tan_{\tau(p)}\pi_{\bar\mu}=
\ker\,(\omega_{\bar\mu})_{\tau(p)}$ and therefore
$v\in\ker\,(\Omega_\mu)_p$ because
$(\Tan_p\tau)^{-1}(\ker\, (\omega_{\bar\mu})_{\tau(p)})=\ker\, (\Omega_\mu)_p$.
But $\ker\,(\Omega_\mu)_p=\ker\,\Tan_p\pi_\mu$, therefore
$u=\Tan_p\pi_\mu(v)=0$, hence $\Tan_{\pi_\mu(p)}\rho$
is injective and, as a consequence, it is an isomorphism.
\item
$\rho$ is a symplectomorphism:

On the one hand
$\Omega_\mu=\pi_\mu^*\hat\Omega$, but on the other hand
$$
\Omega_\mu =
\tau^*\omega_{\bar\mu}=
\tau^*\pi_{\bar\mu}^*\hat\omega=
\pi_\mu^*\rho^*\hat\omega
$$
hence $\pi_\mu^*\hat\Omega=\pi_\mu^*\rho^*\hat\omega$
and, since $\pi_\mu$ is a submersion, we have that
$\hat\Omega=\rho^*\hat\omega$.
\item
The proof for ${\rm H}_\mu=\tau^*{\rm H}_{\bar\mu}$
is like in the last item.
\een
\qed

\subsection{Other reduction procedures:
Coisotropic imbedding {\it plus} symplectic reduction}
\protect\label{coim}

Let $(M,\Omega,{\rm H})$ be a compatible presymplectic dynamical
system and $({\bf M},{\bf \Omega},{\bf H})$ an ambient symplectic system
associated to it (see section \ref{rpds}).
If the presymplectic dynamical system exhibits
non-gauge symmetries as well as gauge symmetries,
under certain hypothesis, both can be removed
applying the standard symplectic reduction procedure of Marsden-Weinstein
to the symplectic dynamical system
$({\bf M},{\bf \Omega},{\bf H})$.
On the other hand, we can apply the presymplectic reduction method
here explained and then we will prove that both procedures also
lead to the same final result.

Let $G$ be a group of symmetries of the presymplectic dynamical system,
${\bf g}$ its Lie algebra and $\tilde{\bf g}\subset\vf_h(M)$ the
corresponding set of fundamental vector fields.
First of all, using the coisotropic imbedding theorem,
it can be proved \cite{CGIR-85} that,
for every presymplectomorphism $\Phi_g\colon (M,\Omega)\to (M,\Omega)$,
there exists a symplectomorphism
${\bf \Phi}_g\colon ({\bf M},{\bf \Omega})\to ({\bf M},{\bf \Omega})$
such that it reduces to $\Phi_g$ on $M$; that is,
${\bf \Phi}_g\circ j_0=j_0\circ\Phi_g$.
Taking this into account, we will assume that:

\begin{assum}
The presymplectic action $\Phi\colon M\times G\to M$ can be
extended to a symplectic action ${\bf \Phi}\colon {\bf M}\times G\to{\bf M}$
which reduces to $\Phi$ on $M\times G$, that is,
${\bf \Phi}_g\circ (j_0\times{\rm Id}_G)=j_0\circ\Phi_g$
(and it is also assumed to be Poissonian, free and proper).
\label{asum4}
\end{assum}

Then, denoting by $\tilde{\bf g}_{\bf M}\subset\vf_h ({\bf M})$
the set of fundamental vector fields for this extended action,
it is obvious that $\tilde{\bf g}_{\bf M}\subset\underline{\vf (M)}$
and then, for every $\tilde\xi'\in\tilde{\bf g}_{\bf M}$,
there exists one $\tilde\xi\in\tilde{\bf g}$ such that
$j_{0_*}\tilde\xi =\tilde\xi'\vert_M$, and conversely.

Let ${\bf J}\colon{\bf M}\to{\bf g}^*$ be
a momentum map associated with the extended action ${\bf \Phi}$.
Once again, the standard symplectic reduction program can be applied:
\bit
\item
First constructing the level sets
$({\bf J}^{-1}(\mu),{\bf \Omega}_\mu,{\bf H}_\mu)$,
for weakly regular values $\mu\in{\bf g}^*$ of ${\bf J}$.
Each one is locally defined in $({\bf M},{\bf \Omega})$
by the constraints $\{{\bf  f}_\xi\}$ which are the
Hamiltonian functions of the vector fields of
$\tilde{\bf g}_{\bf M}$.
\item
Second taking the isotropy group $G_\mu$
and its Lie algebra ${\bf g}_\mu$, for which we have
$\tilde{\bf g}_\mu=\ker\,{\bf \Omega_\mu}$,
and constructing the quotient ${\bf J}^{-1}(\mu)/G_\mu$.
\eit
After that, the corresponding reduced symplectic Hamiltonian system
$({\bf J}^{-1}(\mu)/\ker\,{\bf \Omega}_\mu,\hat{\bf \Omega},\hat{\bf H})$
is free of symmetries.

On the other hand, we can consider the momentum map
${\cal J}\colon M\to{\bf g}^*$ which is {\sl compatible}
with ${\bf J}$. This means that ${\cal J}$ is
induced by ${\bf J}$ on $M$; that is, ${\cal J}:={\bf J}\circ j_0$.
We can then apply the complete presymplectic reduction procedure
constructing the level sets
$({\cal J}^{-1}(\mu),\Omega_\mu,{\rm H}_\mu)$,
for weakly regular values $\mu\in{\bf g}^*$ of ${\cal J}$,
and the quotient ${\cal J}^{-1}(\mu)/G_\mu$.
So we have the (commutative) diagram
$$
\matrix{standard \cr symplectic \cr reduction}
\left\{
\begin{array}{ccc}
({\bf M},{\bf \Omega},{\bf H}) &
\begin{picture}(24,5)(0,0)
\put(24,0){\vector(-1,0){24}}
\put(10,4){\mbox{$j_0$}}
\end{picture}
& (M,\Omega,{\rm H})
\\
\Big\uparrow\ {\bf j}_\mu& & \Big\uparrow\ j_\mu
\\
({\bf J}^{-1}(\mu),{\bf \Omega}_\mu,{\bf H}_\mu) &
\begin{picture}(24,5)(0,0)
\put(24,0){\vector(-1,0){24}}
\put(10,4){\mbox{$j_{0\mu}$}}
\end{picture}
& ({\cal J}^{-1}(\mu),\Omega_\mu,{\rm H}_\mu)
\\
\Big\downarrow\ \pi'_\mu & & \Big\downarrow\ \pi_\mu
\\
({\bf J}^{-1}(\mu)/\ker\,{\bf \Omega}_\mu,\hat{\bf \Omega},\hat{\bf H}) &
\begin{picture}(24,5)(0,0)
\put(24,0){\vector(-1,0){24}}
\put(10,4){\mbox{$\hat\j_0$}}
\end{picture}
& ({\cal J}^{-1}(\mu)/\ker\,\Omega_\mu,\hat\Omega,\hat{\rm H})
\end{array}
\right\}
\matrix{complete \cr presymplectic \cr reduction}
$$
Then, if $\mu\in{\bf g}^*$ is a weakly regular value
for ${\bf J}$ and ${\cal J}$, we wish to compare the sets
$({\bf J}^{-1}(\mu),{\bf \Omega}_\mu,{\bf H}_\mu)$
and $({\bf J}^{-1}(\mu)/\ker\,{\bf \Omega}_\mu,\hat{\bf \Omega},\hat{\bf H})$
with $({\cal J}^{-1}(\mu),\Omega_\mu,{\rm H}_\mu)$
and $({\cal J}^{-1}(\mu)/\ker\,\Omega_\mu,\hat\Omega,\hat{\rm H})$
respectively. So we have:

\begin{teor}
With the conditions stated in the assumption \ref{asum4}:
\ben
\item
$j_{0\mu}({\cal J}^{-1}(\mu))$ is a connected component of ${\bf J}^{-1}(\mu)$,
or a union of connected components of it.
Moreover, $j_{0\mu}^*{\bf  \Omega}_\mu=\Omega_\mu$ and
$j_{0\mu}^*{\bf H}_\mu={\rm H}_\mu$.
\item
$\hat\j_0({\cal J}^{-1}(\mu)/\ker\,\Omega_\mu)$
is a connected component of the quotient
${\bf J}^{-1}(\mu)/\ker\,{\bf \Omega}_\mu$,
or a union of connected components of it.
Moreover, $\hat\j_0^*\hat{\bf \Omega}=\hat\Omega$ and
$\hat\j_0^*\hat{\bf H}=\hat{\rm H}$.
\een
\label{inclusion}
\end{teor}
\proof
Let $\{\xi_i\}$ be a basis of ${\bf g}$ and
${\bf f}_{\xi_i}\in\Cinfty ({\bf M})$
Hamiltonian functions for the corresponding
$\tilde\xi'_i\in\tilde{\bf g}_{\bf M}$.
Let $\{ \alpha^i \}$ be the dual basis in ${\bf g}^*$
and $\mu =\mu_i\alpha^i$ (with $\mu_i\in\Real$)
a weakly regular value for ${\bf J}$ and ${\cal J}$.

First, we are going to prove that, f the submanifold ${\bf J}^{-1}(\mu)\hookrightarrow{\bf M}$
is locally defined by the constraints $\zeta_i:={\bf f}_{\xi_i}-\mu_i$,
then the constraints $j_0^*\zeta_i$ define locally the submanifold
${\cal J}^{-1}(\mu)\hookrightarrow{\bf M}$. In fact;
as ${\cal J}:={\bf J}\circ j_0$ we have that
${\bf J}(x)={\cal J}(x)$, for every $x\in M$, and,
taking into account (\ref{expmm}), this implies that
$j_0^*{\bf f}_{\xi_i}(x)\alpha^i=f_{\xi_i}(x)\alpha^i$
and thus $j_0^*{\bf f}_{\xi_i}=f_{\xi_i}$.
Therefore:
\bit
\item
If $\tilde\xi'_i\not\in\underline{\ker\,\Omega}$
(that is, it generates infinitesimal non-gauge symmetries), then
the corresponding $\tilde\xi_i\in\vf (M)$ is not in $\ker\,\Omega$
and hence
$$
\d j_0^*{\bf  f}_{\xi_i}=j_0^*\d{\bf  f}_{\xi_i}=
j_0^*(\inn(\tilde\xi'_i){\bf \Omega})=
\inn(\tilde\xi_i)\Omega=\d f_{\xi_i}
$$
As $\d f_{\xi_i}\not= 0$ then $j_0^*{\bf f}_{\xi_i}$ is not constant
on $M$, but $j_\mu^*j_0^*{\bf f}_{\xi_i}$ is;
hence ${\bf f}_{\xi_i}$ are constraints for
${\cal J}^{-1}(\mu)\hookrightarrow{\bf M}$
necessarily, but not for $M\hookrightarrow{\bf M}$.

Conversely, every constraint function $\zeta\in\Cinfty (M)$ for
${\cal J}^{-1}(\mu)\hookrightarrow{\bf M}$
can be extended to a function of $\zeta'\in\Cinfty ({\bf M})$
such that its Hamiltonian vector field does not belong to
$\underline{\ker\,\Omega}$.
\item
If $\tilde\xi'_i\in\underline{\ker\,\Omega}$
(that is, it generates infinitesimal gauge symmetries), then
the corresponding $\tilde\xi_i\in\vf (M)$ is in $\ker\,\Omega$ and hence
$$
0=\inn(\tilde\xi_i)\Omega=\inn(\tilde\xi_i)(j_0^*{\bf\Omega})=
j_0^*(\inn(\tilde\xi'_i){\bf \Omega})=j_0^*\d{\bf  f}_{\xi_i}=
\d j_0^*{\bf  f}_{\xi_i}
$$
So ${\bf f}_{\xi_i}$ is constant on $M$ and therefore it
is a constraint for ${\cal J}^{-1}(\mu)\hookrightarrow{\bf M}$,
as well as for $M\hookrightarrow{\bf M}$.

Conversely, for every constraint function
$\phi$ for $M\hookrightarrow{\bf M}$,
its Hamiltonian vector field necessarily belongs to $\underline{\ker\,\Omega}$.
\eit
On the other hand, since $j_0$ is a coisotropic imbedding, we have that
$\dim\, (\ker\,\Omega)=\dim\, {\bf M}-\dim\, M$; and as
$\dim\, ({\bf J}^{-1}(\mu))$ (in ${\bf M}$) is equal to $\dim\,{\bf M}-\dim\,{\bf g}$,
we obtain that
$$
\dim\, ({\cal J}^{-1}(\mu))=\dim\,{\bf M}-(\dim\,{\bf g}-\dim\, (\ker\,\Omega))=
\dim\, M-\dim\,{\bf g}=\dim\, ({\bf J}^{-1}(\mu))
$$
Hence we conclude that $j_{0\mu}({\cal J}^{-1}(\mu))$ is a
submanifold of ${\bf J}^{-1}(\mu)$ and, as both have the same dimension,
we can conclude that $j_{0\mu}({\cal J}^{-1}(\mu))$ is open in ${\bf J}^{-1}(\mu)$.
So, it is a connected component of ${\bf J}^{-1}(\mu)$,
or a union of connected components of it
(remember that both manifolds are closed, since they are defined by constraints).
In addition,
$$
j_{0\mu}^*{\bf\Omega}_\mu=
j_{0\mu}^*{\bf j}_\mu^*{\bf\Omega}=j_\mu^*j_0^*{\bf\Omega}=\Omega_\mu
$$
and, in the same way $j_{0\mu}^*{\bf H}_\mu={\rm H}_\mu$.

Finally, the results for the reduced phase spaces
follow immediately from here.
\qed

A particular case of this reduction procedure
(coisotropic imbedding {\it plus} symplectic reduction)
is when $\ker\,\Omega=\Cinfty (M)\otimes\tilde{\bf g}_M$. Then
$({\bf J}^{-1}(\mu),{\bf \Omega},{\bf H})=(M,\Omega,{\rm H})$
and
$({\bf J}^{-1}(\mu)/\ker\,{\bf \Omega}_\mu,\hat{\bf \Omega},\hat{\bf H})=
(\bar M,\bar\omega,\bar{\rm H})$.
In this case, this method is the {\sl generalized symplectic
reduction} studied in \cite{IM-95} and \cite{MSSV-85}.
(See also \cite{MR-93} for a study on other features on this topic).

The successive steps of the three reduction procedures here analyzed can be
summarized in the following diagram
$$
\begin{array}{ccccc}
& \matrix{coisotropic \cr imbedding} &
& \matrix{gauge \cr reduction} &
\\
\\
({\bf M},{\bf \Omega},{\bf H}) &
\begin{picture}(24,5)(0,0)
\put(24,0){\vector(-1,0){24}}
\put(10,4){\mbox{$j_0$}}
\end{picture}
& (M,\Omega,{\rm H}) &
\begin{picture}(24,5)(0,0)
\put(0,0){\vector(1,0){24}}
\put(10,4){\mbox{$\pi_M$}}
\end{picture}
& (\bar M,\bar\omega,\bar{\rm H})
\\
\Big\uparrow\ {\bf j}_\mu& & \Big\uparrow\ j_\mu & & \Big\uparrow\ j_{\bar\mu}
\\
({\bf J}^{-1}(\mu),{\bf \Omega}_\mu,{\bf H}_\mu) & = &
({\cal J}^{-1}(\mu),\Omega_\mu,{\rm H}_\mu) &
\begin{picture}(24,5)(0,0)
\put(0,0){\vector(1,0){24}}
\put(10,4){\mbox{$\tau$}}
\end{picture}
& (\bar{\cal J}^{-1}(\bar\mu),\omega_{\bar\mu},{\rm H}_{\bar\mu})
\\
\Big\downarrow\ \pi_\mu &
& \Big\downarrow\ \pi_\mu &
& \Big\downarrow\ \pi_{\bar\mu}
\\
({\bf J}^{-1}(\mu)/\ker\,{\bf \Omega}_\mu,\hat{\bf \Omega},\hat{\bf H}) & = &
({\cal J}^{-1}(\mu)/\ker\,\Omega_\mu,\hat\Omega,\hat{\rm H}) &
\begin{picture}(24,5)(0,0)
\put(11,9){\mbox{$\rho$}}
\put(10,0){\mbox{$\simeq$}}
\end{picture}
& (\bar{\cal J}^{-1}(\bar\mu)/\ker\,\omega_{\bar\mu},\hat\omega,\hat{\rm h})
\\
\\
\matrix{standard \cr symplectic \cr reduction} &
& \matrix{complete \cr presymplectic \cr reduction} &
& \matrix{standard \cr symplectic \cr reduction}
\end{array}
$$
(where the equalities mean that the imbeddings of the
manifolds in the center column are (a union of) connected components
of the corresponding manifolds in the left column).

\subsection{Reduction of non-compatible presymplectic dynamical systems
with symmetry}
\protect\label{rpdss}

The concept of group of symmetries can also
be established for a non-compatible presymplectic
dynamical system $(P,\omega ,{\cal H})$ with final compatible system
$(M,\Omega,{\rm H})$ (where $M$ is the final constraint submanifold).
Thus, from the definition \ref{simg} we state:

\begin{definition}
Let $(P,\omega ,{\cal H})$ be
a non-compatible presymplectic dynamical system
with final compatible system $(M,\Omega,{\rm H})$,
$G$ a Lie group and $\Psi\colon G\times P\to P$ an action of $G$ on $P$.
$G$ is said to be a {\rm symmetry group} of this system iff
\ben
\item
$\Psi$ leaves $M$ invariant;
that is, it induces an action $\Phi\colon G\times M\to M$ such that
$\Psi\circ ({\rm Id}_G\times j_M)=j_M\circ\Phi$.
\item
The induced action $\Phi$ is a presymplectic action on $(M,\Omega )$
(which is assumed to be Poissonian, free and proper); that is,
for every $g\in G$,
$$
j_M^*(\Psi_g^*\omega -\omega )=\Phi_g^*\Omega -\Omega =0
$$
(Following a very usual terminology in physics,
we will say that $\Psi$ is a
{\sl weakly presymplectic action} on $(P,\omega,M)$).
\item
For every $g\in G$,
$$
j_M^*(\Psi_g^*{\cal H}-{\cal H})=\Phi_g^*{\rm H}-{\rm H}=0
$$
\een
\label{simgex}
\end{definition}

Of course all the results discussed in the above sections hold
for the compatible dynamical system $(M,\Omega,{\rm H})$
and the action $\Phi$. In particular, let
${\bf g}$ be the Lie algebra of $G$ and
$\tilde{\bf g}_M$ and $\tilde{\bf g}_P$ the sets of
fundamental vector fields on $M$ and $P$
(with respect to the actions $\Phi$ and $\Psi$)  respectively.
Then, for every $\xi\in{\bf g}$, there exist
$\tilde\xi^M\in\tilde{\bf g}_M$ and
$\tilde\xi^P\in\tilde{\bf g}_P$,
with $j_{M_*}\tilde\xi^M=\tilde\xi^P\vert_M$, such that
\beann
j_M^*\Lie(\tilde\xi^P){\cal H}=\Lie(\tilde\xi^M){\rm H}&=&0
\\
j_M^*\Lie(\tilde\xi^P)\omega=\Lie(\tilde\xi^M)\Omega&=&0
\eeann

As the level sets of the momentum map associated to the action
$\Phi$ are submanifolds of $M$, ${\cal J}^{-1}(\mu)\hookrightarrow M$,
we may ask how are they defined as submanifolds of $P$.
We have the following diagram
$$
\begin{array}{ccc}
(P,\omega) &
\begin{picture}(24,5)(0,0)
\put(0,0){\vector(1,0){24}}
\put(10,4){\mbox{$\Psi_g$}}
\end{picture}
& (P,\omega)
\\
j_M\ \Big\uparrow & & \Big\uparrow\ j_M
\\
(M,\Omega) &
\begin{picture}(24,5)(0,0)
\put(0,0){\vector(1,0){24}}
\put(10,4){\mbox{$\Phi_g$}}
\end{picture}
& (M,\Omega)
\\
j_\mu\ \Big\uparrow & & \Big\uparrow\ j_\mu
\\
({\cal J}^{-1}(\mu),\Omega_\mu) & & ({\cal J}^{-1}(\mu),\Omega_\mu)
\end{array}
$$
Taking into account the above discussion, we have that
the constraint functions $\zeta^M_i\in\Cinfty (M)$
defining ${\cal J}^{-1}(\mu)$ in $M$ can be extended to
$P$ as functions $\zeta^P_i\in\Cinfty (P)$
such that $j_M^*\zeta^P_i=\zeta^M_i$ since,
if $\{\xi_i\}$ is a basis of ${\bf g}$, we have
$$
\d\zeta^M_i:=\inn (\tilde\xi^M_i)\Omega =j_M^*(\inn (\tilde\xi^P_i)\omega)=
j_M^*\d\zeta^P_i=\d (j_M^*\zeta^P_i)
$$
Hence the submanifolds ${\cal J}^{-1}(\mu)\hookrightarrow P$
are defined in $P$ by these constraints $\zeta^P$
together with the constraints $\eta$ defining $M$ in $P$.

As a particular case of special interest we have:

\begin{prop}
With the conditions of definition \ref{simgex},
and assuming the following hypothesis:
\ben
\item
Assumption \ref{asum2} holds for $(M,\Omega)$.
\item
There is a basis of constraint functions $\{\eta_i\}$
defining $M$ in $P$ made of presymplectic Hamiltonian functions
in $(P,\omega)$.
\item
The presymplectic Hamiltonian vector fields $X_{\eta_i}\in\vf_h(P)$
associated to these constraints are tangent to $M$.
\een
Then the momentum map ${\cal J}\colon M\to {\bf g}^*$ can be extended
to a map ${\rm J}\colon P\to{\bf g}^*$ such that
${\cal J}={\rm J}\circ j_M$ and ${\cal J}^{-1}(\mu)={\rm J}^{-1}(\mu)$.
\label{mapexpre}
\end{prop}
\proof
First of all, from items 2 and 3 we have that $X_{\eta_i}$
expand locally the set $\underline{\ker\,\Omega}$
\cite{Go-79}, \cite{BK-86} and, from item 1, we have also that $X_{\eta_i}\in\tilde{\bf g}$.
Then, the map ${\rm J}\colon P\to{\bf g}^*$ is defined in the
following way: for every $\xi\in{\bf g}$ let
$f^M_{\xi}\in\Cinfty (M)$ be the presymplectic Hamiltonian function
of the fundamental vector field $\tilde\xi^M\in\vf (M)$
and $f^P_{\xi}\in\Cinfty (P)$ its extension to $P$;
then, for every $p\in P$,
$$
({\rm J}(p))(\xi):=f_{\xi}^P(p)
$$
Observe that if $\tilde\xi^M\in\ker\,\Omega$ then
$f^M_{\xi}$ is constant and therefore
$f^P_{\xi}\equiv\eta$ is a constraint for $M$ in $P$;
whereas if $\tilde\xi^M\not\in\ker\,\Omega$ then
$f^M_{\xi}$ is a constraint for ${\cal J}^{-1}(\mu)$ in $M$.
Thus ${\cal J}={\rm J}\circ j_M$ and ${\cal J}^{-1}(\mu)={\rm J}^{-1}(\mu)$
(see the proof of theorem \ref{inclusion}).
\qed

It is important to point out that, in general,
${\rm J}$ is not strictly speaking a momentum map,
because the action $\Psi$ is not necessarily
presymplectic for $(P,\omega)$.

In any case, if the assumption \ref{asum2} is assumed for $(M,\Omega)$,
the reduction program follows in the same way
as in the case of compatible presymplectic dynamical systems.

\section{Examples}

\subsection{Reduction of non-autonomous systems with symmetry in
the presymplectic formulation}
\protect\label{rnass}

Non-autonomous dynamical systems can be geometrically treated in
several ways (see, for instance,
\cite{AM-78}, \cite{CPT-84}, \cite{EMR-91}, \cite{HL-84},
\cite{Ku-84}, \cite{Ra-91} for a review on these formulations).
Reduction of time-dependent systems with symmetry can be achieved
by using the {\sl extended } or {\sl symplectic formulation},
and then by using, then, the usual reduction theory for symplectic
systems with symmetry
(see, for instance, \cite{CCCI-86} and \cite{LS-93}).
Nevertheless, we will use reduction for presymplectic systems,
(whose features we have just presented) because it has
some advantages in relation to the symplectic case;
for instance, {\sl singular time-dependent systems} with
symmetry can be treated in this formulation in a very natural way.

Thus we need to use the {\sl presymplectic formulation}
of non-autonomous systems \cite{CGIR-87}, \cite{EMR-91}.
The main characteristics of this formulation are the following:
the dynamics takes place in a differentiable manifold $P\times\Real$,
where $(P,\omega_P)$ is either a symplectic manifold,
if the non-autonomous system is {\sl regular}, or a presymplectic one,
if it is {\sl singular}; and we have the natural projections
\beq
\tau\colon P\times\Real\to P
\quad ; \quad
t\colon P\times\Real\to \Real
\label{projs}
\eeq
The dynamical information is entirely contained
in a function $h\in\Cinfty (P\times\Real )$.
Then, $P\times\Real$ is endowed with the following presymplectic structure
$$
\Omega_h :=\tau^*\omega_P+\d h\wedge\d t
$$
which is exact if, and only if, so is $\omega_P$.
So, we have the dynamics fully included in the geometry
and, therefore, we can obtain the equations of motion stating that
$(P\times\Real ,\Omega_h ,0)$ is a presymplectic Hamiltonian system;
that is, in the presymplectic equations of motion
$\inn (X)\Omega_h =\d{\rm H}$,
we take ${\rm H}=0$. Then the equations of motion are reduced to
$$
\inn (X)\Omega_h =0
\ , \
X\in\vf (P\times\Real)
$$
Since $\d{\rm H}=0$, these equations are compatible in $P\times\Real$
and, consequently, there exists solution $X\in\ker\,\Omega_h$.
On the other hand, if we want to yield the time-reparametrization $t=s$
we must add the equation \dst\inn (X)\d t= 1\) ;
(however, other possible reparametrizations
having physical sense are also possible \cite{ACI-83}, \cite{CIL-88}).

We differentiate the following situations:
\bit
\item
The {\sl Lagrangian formalism of non-autonomous systems}:

$P$ is the tangent bundle $\Tan Q$ of the configuration space $Q$.
Then, given a {\sl time-dependent Lagrangian function}
$\Lag\in\Cinfty (\Tan Q\times\Real )$, using the extensions
to $\Tan Q\times\Real$ of the natural geometric structures
in the tangent bundle (the {\sl vertical endomorphism} and the
{\sl Liouville's vector field}), we can construct the exact
form $\omega_{\Lag}\in\df^2(\Tan Q\times\Real)$,
which plays the role of the form $\tau^*\omega_P$,
and the {\sl energy Lagrangian function}
${\rm E}_{\Lag}\in\Cinfty (\Tan Q)$ which plays the role of $h$
in this formalism. Then
$$
\Omega_h\equiv\Omega_{\Lag}=\omega_{\Lag}+\d{\rm E}_{\Lag}\wedge\d t
$$
(see \cite{EMR-91} and \cite{EMR-95}
for a discussion on the construction of these elements).

If the system is {\sl regular}, then
the form $\omega_{\Lag}$ is symplectic.
If the system is {\sl singular} then $\omega_{\Lag}$ is a presymplectic form.
\item
The {\sl Hamiltonian formalism of non-autonomous systems}:

If the system is not {\sl singular}, then
$(P,\omega_P)$ is a symplectic manifold,
$P$ being the cotangent bundle $\Tan^*Q$ of the configuration space $Q$
(if it is {\sl hiper-regular})
or an open submanifold of it (if it is {\sl regular})
and $\omega_P\equiv\omega\in\df^2(\Tan^*Q)$ being its natural canonical form,
which is an exact form. Then,
$h\in\Cinfty (\Tan^*Q)$ is the {\sl time-dependent Hamiltonian function}.

If the system is {\sl singular}, then
$(P,\omega_P)$ is a presymplectic manifold,
$j\colon P\hookrightarrow\Tan^*Q$ being a submanifold
of the cotangent bundle $\Tan^*Q$ of the configuration space $Q$
(really it is the image of $\Tan Q$ by the {\sl Legendre transformation})
and $\omega_P=j^*\omega$.
Then, $h\in\Cinfty (\Tan^*Q)$ is called the
{\sl canonical time-dependent Hamiltonian function}.
\eit

Concerning the study of symmetries, time-dependent dynamical systems display
some particular characteristics which are interesting to point out.
Thus, from the geometrical and the dynamical point of view,
a natural way of defining the concept of symmetry is the following:

\begin{definition}
Let $G$ be a Lie group, $(P\times\Real ,\Omega_h)$
a non-autonomous system
and $\Phi\colon G\times (P\times\Real )\to P\times\Real$
an action of $G$ on $P\times\Real$.
$G$ is said to be a {\rm group of standard symmetries} of this system iff,
for every $g\in G$,
\ben
\item
$\Phi_g$ preserves the forms $\tau^*\omega_P$ and $\d t$; that is,
$$
\Phi_g^*\tau^*\omega_P =\tau^*\omega_P
\quad ;\quad
\Phi_g^*\d t =\d t
$$
\item
$\Phi_g$ preserves the dynamical function $h$; that is,
$\Phi_g^*h=h$.
\een
The diffeomorphisms $\Phi_g$ are called {\rm standard symmetries} of the system.
\label{stansym}
\end{definition}

The first part of this definition is equivalent to that of
{\sl cosymplectic action} introduced in
\cite{LS-93} and agrees also with the concept of
{\sl standard canonical transformation} for time-dependent
Hamiltonian systems, which other authors have previously introduced
see \cite{ACI-83}, \cite{CIL-88}.

As immediate consequences of this definition we have that:
\bit
\item
If $G$ is a group of standard symmetries of the
non-autonomous system $(P\times\Real ,\Omega_h)$ then,
for every $g\in G$, every standard symmetry $\Phi_g$ preserves
the form $\Omega_h$; that is,
$\Phi_g^*\Omega_h =\Omega_h$.
\item
$G$ is a group of standard symmetries of the non-autonomous system
$(P\times\Real ,\Omega_h)$ if, and only if,
the following three conditions hold for every $\xi\in{\bf g}$:
$$
{\rm (1)}\qquad
\Lie(\tilde\xi )(\tau^*\omega_P)=0
\quad , \qquad {\rm (2)}\qquad
\Lie(\tilde\xi )\d t=0
\quad , \qquad{\rm (3)}\qquad
\Lie(\tilde\xi )h=0
$$
\item
If $G$ is a group of standard symmetries for the non-autonomous system
$(P\times\Real ,\Omega_h)$ then it is also a symmetry group
for the presymplectic Hamiltonian system $(P\times\Real ,\Omega_h,0)$
(in the sense of definition \ref{simg}).
\eit

At this point, reduction of non-autonomous dynamical systems
with symmetry (both in the Lagrangian or in the Hamiltonian
formalism) is merely a direct application of the
considerations we made above in order to reduce
presymplectic systems with symmetry.

\subsection{Autonomous dynamical systems}
\protect\label{ads}

As an example, we are going to analyze the
time-independent dynamical systems as the particular case
of non-autonomous regular systems which are invariant under time-translations.
This study is identical for the Lagrangian and the Hamiltonian
formalism and we will do it in general.

Let $(P\times\Real ,\Omega_h)$ be a
non-autonomous regular dynamical system
(then $P$ is either $\Tan^*Q$ or $\Tan Q$ and ${\rm dim}\, P=2r$).
$G$ is the group of translations in time.
The action $\Phi\colon G\times (P\times\Real )\to P\times\Real$ is
effective, free and proper.
The real Lie algebra ${\bf g}$ is spanned by the vector field
\dst\xi\equiv\derpar{}{t}\) and hence ${\bf g}^*=\{\d t\}$.
Thus, the set of fundamental vector fields $\tilde{\bf g}$ is generated by
the vector field \dst\tilde\xi\equiv\derpar{}{t}\) .

Suppose that the dynamical function $h$ is time-independent,
that is,
$$
\Lie\left(\derpar{}{t}\right)h=\derpar{h}{t}=0
$$
It is evident that this action verifies the conditions of
definition \ref{stansym},
and hence $G$ is a symmetry group for this system.
Then the action is presymplectic:
\dst\Lie\left(\derpar{}{t}\right)\Omega_h=0\) , and,
in addition, it is strongly presymplectic since
the fundamental vector field is Hamiltonian
(it is in fact an exact presymplectic action) and
$$
\inn\left(\derpar{}{t}\right)\Omega_h=\d h
$$

In this way the momentum map is given by
$$
({\cal J}(x))\left(\derpar{}{t}\right):=h(x)
\quad \mbox{\rm (for every $x\in P\times\Real$)}
$$
and the level sets of this map, for every weakly regular value
$\mu =\mu_0\d t\in{\bf g}^*$, are
$$
{\cal J}^{-1}(\mu ):=\{ x\in P\times\Real\ \vert\ h(x)=\mu_0\}
$$
(they are defined by the constraints $\zeta :=h-\mu_0$
and hence \dst\derpar{}{t}\) is tangent to all of them).
They are the hypersurfaces of constant energy in $P\times\Real$.

In each one, we have the presymplectic Hamiltonian system
$({\cal J}^{-1}(\mu ),\Omega_\mu ,0)$, where
$\Omega_\mu :=j_\mu^*\Omega_h=j_\mu^*\tau^*\Omega_P$.
Notice that, even though ${\rm dim}\,{\cal J}^{-1}(\mu )$ is even,
$\Omega_\mu$ is presymplectic since
\dst\ker\Omega_\mu=\left\{\derpar{}{t},X_\mu\right\}\) ,
where $X_\mu\in\vf ({\cal J}^{-1}(\mu ))$ is the
solution of the dynamical equation
$\inn (X_\mu)\Omega_\mu =0$.
So, in this case, since $G=G_\mu$, $\tilde{\bf g}_\mu$ is generated by
the vector field \dst\tilde\xi\equiv\derpar{}{t}\) and
we have that $\tilde{\bf g}_\mu\subset\ker\,\Omega_\mu$.
Therefore, applying the reduction theorems,
we have the same situation as the first item in theorem \ref{MWt},
and hence this presymplectic system reduces to another one
$({\cal J}^{-1}(\mu )/G,\hat\Omega ,0)$.
This is a $(2r-1)$-dimensional differentiable manifold
(and then $\hat\Omega$ is a presymplectic form with
${\rm rank}\,\hat\Omega =2r-2$)
where the global coordinate $t$ is avoided.
The evolution equations are
$$
\inn (\hat X)\hat\Omega =0
\quad ,\quad \tilde X\in\vf({\cal J}^{-1}(\mu )/G)
$$
Observe that the main advantage of this reduction procedure
is that, in addition to eliminating the ignorable time-coordinate,
it already gives the solution of dynamics directly
on the corresponding hypersurface of constant energy.
This is an advance in relation to the use of the
symplectic reduction procedure of Marsden and Weinstein,
applied for treating this same example but starting from
the extended symplectic formalism of the non-autonomous systems.
In this case, even though the reduced dynamical system is regular
(and then symplectic), the symplectic reduction procedure
removes time only (from the initial time-dependent system),
but it does not give the dynamics on the constant-energy hypersurfaces
which is obtained by projection, that is, after another step not included
in the reduction procedure (see \cite{LS-93}).

Nevertheless, a further reduction could be made by the residual part
of $\ker\Omega_\mu$ (that is those one generated by $X_\mu$) or,
what is equivalent, make the reduction of
the presymplectic Hamiltonian system
$({\cal J}^{-1}(\mu ),\Omega_\mu ,0)$
by $\ker\,\Omega_\mu$. In this way we would have the situation of
the second item in theorem \ref{MWt},
and hence this presymplectic system reduces to
$({\cal J}^{-1}(\mu )/\ker\,\Omega_\mu,\hat\Omega',0)$,
where ${\cal J}^{-1}(\mu )/\ker\,\Omega_\mu$
is a $(2r-2)$-dimensional differentiable manifold
and $\hat\Omega'$ is a symplectic form.
As a consequence, there is no dynamics in this reduced system;
that is, the orbit space is made of the dynamical trajectories
of the initial time-independent dynamical system
(for a fixed constant value of the energy).

\subsection{A mechanical model of field theories: description}
\protect\label{mmft1}

The following example we study is based
in a {\sl mechanical model of field theories}
(coupled to external fields) due to {\it Capri} and {\it Kobayashi}
\cite{CK-82}, \cite{CK-87}. See also \cite{Cox-92}
for a deeper analysis.

The general form of the Lagrangian of the system is
$$
\Lag = \dot\psi^{*a}m_{ab}\dot\psi^b+\dot\psi^{*a}c_{ab}\psi^b-
\psi^{*a}\bar c_{ab}\dot\psi^b-\psi^{*a}r_{ab}\psi^b
$$
where:
\bit
\item[-]
$\psi^a,\psi^{*b}$ ($a,b=1,\ldots ,n$) are scalar (complex) fields.
In this mechanical model they will be interpreted as independent
``coordinates'' of certain $2n$-dimensional configuration space $Q$.
\item[-]
$m_{ab},c_{ab},\bar c_{ab},r_{ab}$ are (time-independent)
functional coefficients such that, in order $\Lag$ to be real,
the matrices $m_{ab},r_{ab}$ are hermitian and
$\bar c^*_{ab}=-c_{ba}$.
In particular, if ${\rm rank}\,(m_{ab})<n$, the Lagrangian is
singular and this is the case of greatest interest to us.
\eit
Lagrangians of this kind enables us to describe some relativistic
bosonic field theories (after a {\sl (3+1)-decomposition}),
where the eventual coupling to external fields is tucked away in the
coefficients $c_{ab}$ and $r_{ab}$.

In order to make the example more pedagogical, we will analyze
the following simple case:
$a,b=1,2,3$ and
$$
m_{ab}=\left(\matrix{0 & 0 & 0\cr 0 & m_2 & 0 \cr 0 & 0 & m_3 \cr}\right)
\qquad
c_{ab}=\bar c_{ab}=
\frac{1}{2}\left(\matrix{0 & 0 & 0\cr 0 & i & 0 \cr 0 & 0 & i \cr}\right)
\qquad
r_{ab}=\left(\matrix{1 & 0 & 0\cr 0 & 1 & 0 \cr 0 & 0 & 1 \cr}\right)
$$
We can write the Lagrangian in its real form by the change
$$
\begin{array}{ccccc}
\psi^1=x^1+iy^1 & \quad , \quad &
\psi^2=x^2+iy^2 & \quad , \quad &
\psi^3=x^3+iy^3
\\
\dot\psi^1=u^1+iv^1 & \quad , \quad &
\dot\psi^2=u^2+iv^2 & \quad , \quad &
\dot\psi^3=u^3+iv^3
\\
\psi^{*1}=x^1-iy^1 & \quad , \quad &
\psi^{*2}=x^2-iy^2 & \quad , \quad &
\psi^{*3}=x^3-iy^3
\\
\dot\psi^{*1}=u^1-iv^1 & \quad , \quad &
\dot\psi^{*2}=u^2-iv^2 & \quad , \quad &
\dot\psi^{*3}=u^3-iv^3
\end{array}
$$
and hence
\beann
\Lag&=&
m_2((u^2)^2+(v^2)^2)+m_3((u^3)^2+(v^3)^2)+v^2x^2-u^2y^2+v^3x^3-u^3y^3
\\ & &
-(x^1)^2-(y^1)^2-(x^2)^2-(y^2)^2-(x^3)^2-(y^3)^2
\eeann
Here, the configuration space is taken to be $Q=\Real^6$ with
local coordinates $(x^i,y^i)$ ($i=1,2,3$)
and $\Tan Q\simeq\Real^{12}$ with
local coordinates $(x^i,y^i;u^i,v^i)$,
where $u^i,v^i$ denote the generalized velocities
corresponding to $x^i,y^i$ respectively.
Using the canonical structures of the tangent bundle $\Tan Q$,
the {\sl Lagrangian 2-form} and the {\sl energy Lagrangian function}
are constructed:
\beann
\omega_{\Lag}&=&
2[m_2(\d x^2\wedge\d u^2+\d y^2\wedge\d v^2)+
m_3(\d x^3\wedge\d u^3+\d y^3\wedge\d v^3)+
\d x^2\wedge\d y^2+\d x^3\wedge\d y^3]
\\
{\rm E}_{\Lag}&=&
m_2((u^2)^2+(v^2)^2)+m_3((u^3)^2+(v^3)^2)+
(x^1)^2+(y^1)^2+(x^2)^2+(y^2)^2+(x^3)^2+(y^3)^2
\eeann
The system is singular since $\omega_{\Lag}$ is presymplectic and
$$
\ker\,\omega_{\Lag}\equiv
\left\{\derpar{}{x^1},\derpar{}{y^1},\derpar{}{u^1},\derpar{}{v^1}\right\}
$$
$(\Tan Q,\omega_{\Lag},{\rm E}_{\Lag})$ is a presymplectic
dynamical system which is not compatible since
$$
\inn\left(\derpar{}{x^1}\right)\d{\rm E}_{\Lag}=2x^1\not= 0
\quad ,\quad
\inn\left(\derpar{}{y^1}\right)\d{\rm E}_{\Lag}=2y^1\not= 0
$$
So the constraints
$$
\eta_1:=x^1=0 \quad ,\quad \eta_2:=y^1=0
$$
define locally a submanifold $j_M\colon M\hookrightarrow\Tan Q$
where the vector fields which are solutions of the dynamical equation
\beq
(\inn (X)\omega_{\Lag}-\d{\rm E}_{\Lag})\vert_M=0
\label{dineq}
\eeq
are the following
\bea
X\vert_M&=&
f^1\derpar{}{x^1}+u^2\derpar{}{x^2}+u^3\derpar{}{x^3}+
g^1\derpar{}{y^1}+v^2\derpar{}{y^2}+v^3\derpar{}{y^3}+
F^1\derpar{}{u^1}-\frac{1}{m_2}(v^2+x^2)\derpar{}{u^2}
\nonumber
\\
& & -\frac{1}{m_3}(v^3+x^3)\derpar{}{u^3}+G^1\derpar{}{v^1}+
\frac{1}{m_2}(u^2-y^2)\derpar{}{v^2}+\frac{1}{m_3}(u^3-y^3)\derpar{}{v^3}
\label{solution}
\eea
where $f^1,g^1,F^1,G^1$ are arbitrary functions.
Now we consider two options:
\ben
\item
If we look for solutions of the dynamics which are
{\sl second order differential equations} (SODE) then,
in this case, we obtain such a solution taking the first two
arbitrary functions to be $f^1=u^1$ and $g^1=v^1$.
Therefore the stability of this vector field on the constraints
$\eta_1,\eta_2$ originates two new constraints
(which are called {\sl non-dynamical constraints} following
the terminology of \cite{MR-92})
$$
\chi_1:=u^1=0 \quad ,\quad \chi_2:=v^1=0
$$
which, joined to the above ones $\eta_1,\eta_2$,
define locally the submanifold $j_S\colon S\hookrightarrow\Tan Q$.
Finally, the stability of the SODE vector field on the last constraints
fixes the value of the remaining arbitrary functions to be
$F^1=0$, $G^1=0$. So the final constraint submanifold is $S$
(local coordinates are $(x^2,x^3,y^2,y^3,u^2,u^3,v^2,v^3)$ and
\beann
\Omega_S:=j_S^*\omega_{\Lag}&=&
2[m_2(\d x^2\wedge\d u^2+\d y^2\wedge\d v^2)+
m_3(\d x^3\wedge\d u^3+\d y^3\wedge\d v^3)
\\ & &
+\d x^2\wedge\d y^2+\d x^3\wedge\d y^3]
\\
{\rm E}_S:=j_S^*{\rm E}_{\Lag}&=&
m_2((u^2)^2+(v^2)^2)+m_3((u^3)^2+(v^3)^2)+(x^2)^2+(y^2)^2+(x^3)^2+(y^3)^2
\eeann
Observe that, in this example, $(S,\Omega_S)$ is a symplectic manifold.
The SODE vector field tangent to $S$
being the (unique) solution of the dynamical equation
$$
(\inn (X)\omega_{\Lag}-\d{\rm E}_{\Lag})\vert_S=0
$$
is then
\beann
X\vert_S&=&
u^2\derpar{}{x^2}+u^3\derpar{}{x^3}+v^2\derpar{}{y^2}+v^3\derpar{}{y^3}
-\frac{1}{m_2}(v^2+x^2)\derpar{}{u^2}
\\ & &
-\frac{1}{m_3}(v^3+x^3)\derpar{}{u^3}+
\frac{1}{m_2}(u^2-y^2)\derpar{}{v^2}+\frac{1}{m_3}(u^3-y^3)\derpar{}{v^3}
\eeann
(See also \cite{Cox-92} for a more detailed discussion on
this analysis).
\item
If we look for solutions of the dynamics which are not
SODE, then the stability of (\ref{solution}) on the constraints
$\eta_1,\eta_2$ fixes the value of the first two arbitrary functions to be
$f^1=0$, $g^1=0$. So the final constraint submanifold is $M$
(local coordinates are $(x^2,x^3,y^2,y^3,u^1,u^2,u^3,v^1,v^2,v^3)$ and
the coordinate expressions of
$\Omega_M:=j_M^*\omega_{\Lag}$ and ${\rm E}_M:=j_M^*{\rm E}_{\Lag}$
are the same as for $\Omega_S$ and ${\rm E}_S$ respectively.
Hence $(M,\Omega_M)$ is a presymplectic manifold with
$$
\ker\,\Omega_M\equiv\left\{\derpar{}{u^1},\derpar{}{v^1}\right\}
$$
and the vector fields tangent to $M$
being solutions of the dynamical equation (\ref{dineq}) are
\beann
X\vert_M&=&
u^2\derpar{}{x^2}+u^3\derpar{}{x^3}+v^2\derpar{}{y^2}+v^3\derpar{}{y^3}+
F^1\derpar{}{u^1}-\frac{1}{m_2}(v^2+x^2)\derpar{}{u^2}
\\ & &
-\frac{1}{m_3}(v^3+x^3)\derpar{}{u^3}+G^1\derpar{}{v^1}+
\frac{1}{m_2}(u^2-y^2)\derpar{}{v^2}+\frac{1}{m_3}(u^3-y^3)\derpar{}{v^3}
\eeann
\een

\subsection{A mechanical model of field theories: symmetries and reduction}
\protect\label{mmft2}

Next we are going to study the symmetries of the systems,
splitting the two cases considered in the above section;
that is, we will apply the reduction procedure to
the compatible dynamical systems
$(S,\Omega_S,{\rm E}_S)$ and $(M,\Omega_M,{\rm E}_M)$.

Both of them exhibit some non-gauge rigid symmetries which are rotations
on $Q$ whose infinitesimal generators are the following vector fields in $Q$
$$
x^2\derpar{}{y^2}-y^2\derpar{}{x^2}
\quad , \quad
x^3\derpar{}{y^3}-y^3\derpar{}{x^3}
$$
and whose canonical liftings to $\Tan Q$ give the following
fundamental vector fields
\beann
\tilde\xi_1 &=&
x^2\derpar{}{y^2}-y^2\derpar{}{x^2}+u^2\derpar{}{v^2}-v^2\derpar{}{u^2}
\\
\tilde\xi_2 &=&
x^3\derpar{}{y^3}-y^3\derpar{}{x^3}+u^3\derpar{}{v^3}-v^3\derpar{}{u^3}
\eeann
In fact, these vector fields generate infinitesimal symmetries
for these presymplectic systems because
both of them are tangent to $S$ and $M$ and they satisfy that
\beann
j_M^*\Lie(\tilde\xi_k){\rm E}_{\Lag}=0=j_S^*\Lie(\tilde\xi_k){\rm E}_{\Lag}
\\
j_M^*\Lie(\tilde\xi_k)\omega_{\Lag}=0=j_S^*\Lie(\tilde\xi_k)\omega_{\Lag}
\eeann
since $\Lie(\tilde\xi_k){\rm E}_{\Lag}=0$ ($k=1,2$) and
$\Lie(\tilde\xi_k)\omega_{\Lag}=0$,
so both of them are presymplectic Hamiltonian vector fields
in $(S,\Omega_S,{\rm E}_S)$ and $(M,\Omega_M,{\rm E}_M)$.

\ben
\item
{\bf Reduction of the system $(S,\Omega_S,{\rm E}_S)$}.

Since $(S,\Omega_S)$ is a symplectic manifold, there are no gauge symmetries
and the only symmetries to be taken into account are the rigid ones
which have just been introduced.
Denoting by $G$ the corresponding group and by ${\bf g}$
its Lie algebra, then $\tilde{\bf g}\equiv (\tilde\xi^1,\tilde\xi_2)$.
The action considered is in fact strongly presymplectic, since
it is an exact action in relation to the 1-form
$$
j_S^*\theta_{\Lag}=
(2m_2u^2+y^2)\d x^2+(2m_2v^2-x^2)\d y^2+(2m_3u^3-y^3)\d x^3+(2m_3v^3+x^3)\d y^3
$$
The presymplectic Hamiltonian functions
of $\tilde\xi^1$ and $\tilde\xi^2$ in $(S,\Omega_S)$ are
$$
f_{\xi_1}=2m_2(x^2v^2-y^2u^2)-(x^2)^2-(y^2)^2
\quad , \quad
f_{\xi_2}=2m_3(x^3v^3-y^3u^3)-(x^3)^2-(y^3)^2
$$
So a momentum map ${\cal J}_S$ can be defined for this action and,
taking into account the discussion in the section \ref{rpdss},
for every weakly regular value
$\mu\equiv (\mu_1,\mu_2)\in{\bf g}^*$,
its level sets ${\cal J}_S^{-1}(\mu)$ are defined
as submanifolds of $\Tan Q$ by the constraints
\beann
&\eta_1:=x^1=0 \quad ,\quad \eta_2:=y^1=0
\quad ,\quad
\chi_1:=u^1=0 \quad ,\quad \chi_2:=v^1=0&
\\
&f_{\xi_1}:=2m_2(x^2v^2-y^2u^2)-(x^2)^2-(y^2)^2=\mu_1&
\\
&f_{\xi_2}:=2m_3(x^3v^3-y^3u^3)-(x^3)^2-(y^3)^2=\mu_2&
\eeann
The submanifolds $({\cal J}_S^{-1}(\mu),\Omega_{S_\mu})$
are presymplectic and 6-dimensional.
Next, the final step of the reduction procedure leads to the
4-dimensional quotient manifolds
$({\cal J}_S^{-1}(\mu)/\ker\,\Omega_{S_\mu},\hat\Omega_S)$.
\item
{\bf Reduction of the system $(M,\Omega_M,{\rm E}_M)$}.

This compatible presymplectic system
exhibits the above rigid symmetries as well as gauge symmetries,
which are infinitesimally generated by the fundamental vector fields
$$
\tilde\xi_3=\derpar{}{u_1} \quad ,\quad \tilde\xi_4=\derpar{}{v_1}
$$
(which generate $\ker\,\Omega_M$).
Let $G$ be the group of all these symmetries and ${\bf g}$ its Lie algebra,
then $\tilde{\bf g}\equiv (\tilde\xi_1,\tilde\xi_2,\tilde\xi_3,\tilde\xi_4)$.
The action considered is in fact also strongly presymplectic, since
it is an exact action in relation to the 1-form $j_M^*\theta_{\Lag}$.
The presymplectic Hamiltonian functions
of the fundamental vector fields in $(S,\Omega_S)$ are
\beann
f_{\xi_1}:=2m_2(x^2v^2-y^2u^2)-(x^2)^2-(y^2)^2
&\quad , \quad&
f_{\xi_2}:=2m_3(x^3v^3-y^3u^3)-(x^3)^2-(y^3)^2
\\
f_{\xi_3}:=0 &\quad , \quad& f_{\xi_4}:=0
\eeann
where the constant value of $f_{\xi_3}$ and $f_{\xi_4}$
equal to 0 is just a possible choice for the
constant Hamiltonian functions corresponding to the vector fields
$\tilde\xi_3$ and $\tilde\xi_4$ respectively.
So a momentum map ${\cal J}_M$ can be defined for this action and,
taking into account the discussion in the section \ref{rpdss},
for weakly regular values $\mu\equiv (\mu_1,\mu_2,0,0)\in{\bf g}^*$,
its level sets ${\cal J}_M^{-1}(\mu)$ are defined
as submanifolds of $\Tan Q$ by the constraints
\beann
&f_{\xi_1}:=2m_2(x^2v^2-y^2u^2)-(x^2)^2-(y^2)^2=\mu_1&
\\
&f_{\xi_2}:=2m_3(x^3v^3-y^3u^3)-(x^3)^2-(y^3)^2=\mu_2&
\\
&\eta_1:=x^1=0 \quad ,\quad \eta_2:=y^1=0&
\eeann
(Observe that $\eta_1:=x^1$ and $\eta_2:=y^1$
are the presymplectic Hamiltonian functions
of $\tilde\xi_3$ and $\tilde\xi_4$ in $(\Tan Q,\Omega_{\Lag})$, respectively).
Now, the submanifolds $({\cal J}_M^{-1}(\mu),\Omega_{M_\mu})$
are presymplectic and 8-dimensional and the final quotient manifolds
$({\cal J}_M^{-1}(\mu)/\ker\,\Omega_{M_\mu},\hat\Omega_M)$
are 4-dimensional.

Nevertheless, this quotient manifold is locally symplectomorphic to
$({\cal J}_S^{-1}(\mu)/\ker\,\Omega_{S_\mu},\hat\Omega_S)$.
In fact; instead of using the complete presymplectic reduction,
we can apply to the system $(M,\Omega_M,{\rm E}_M)$,
first, gauge reduction and, afterwards, standard symplectic reduction,
then obtaining a quotient manifold which is symplectomorphic to
$({\cal J}_M^{-1}(\mu)/\ker\,\Omega_{M_\mu},\hat\Omega_M)$
(see section \ref{comp1}) and locally symplectomorphic to
$({\cal J}_S^{-1}(\mu)/\ker\,\Omega_{S_\mu},\hat\Omega_S)$
(obviously).
\een

{\bf Comment}:

In this example we have just shown that whether or not
non-dynamical constraints (that is, those arising in the
stabilization algorithm from demanding that the vector field solution
of the Lagrange equations to be a SODE)
are taken into account in the reduction procedure is irrelevant,
since, in any case, we obtain the same quotient manifold.

In reality this must be a general property. In fact:
let $(\Tan Q,\omega_{\Lag},{\rm E}_{\Lag})$ be a singular
(but {\sl almost-regular} \cite{Go-79}, \cite{MR-92})
Lagrangian system, $(M,\Omega_M)$ the final constraint submanifold
when the {\sl SODE}-condition is not considered
and $(S,\Omega_S)$ the final constraint submanifold
when the {\sl SODE}-condition is considered,
such that we have a group of rigid symmetries both for
$(M,\Omega_M,{\rm E}_M)$ and $(S,\Omega_S,{\rm E}_S)$.
Then, if the assumption \ref{asum2} holds for
$(M,\Omega_M)$ and $(S,\Omega_S)$,
the complete presymplectic reduction procedure
leads to the same reduced system for both systems.

The reason for this feature lies in the following facts:
As is proved in \cite{MR-92},
the non-dynamical constraints defining $S$ in $M$
remove degrees of freedom in the leaves of the foliation
generated by the vertical part of $\ker\,\Omega_{\Lag}$
which, on its turn, is included in $\underline{\ker\,\Omega_M}$.
As a consequence, it is also proved that
$\underline{\ker\,\Omega_S}\subset\underline{\ker\,\Omega_M}$.
But, as assumption 2 holds, the final quotient
in the complete presymplectic reduction
is made by a foliation whose leaves contain
those of $\ker\,\Omega_M$ or $\ker\,\Omega_S$ respectively
in each case. Therefore, when the reduction is made for
$(M,\Omega_M)$, the degrees of freedom in the leaves of the foliation
generated by the vertical part of $\ker\,\Omega_{\Lag}$
are removed in the final quotient. However,
when the reduction is made for $(S,\Omega_S)$,
these degrees of freedom have been previously removed.

\subsection{The conformal particle}
\protect\label{cp}

Finally, we consider the system of a
{\sl massless relativistic particle with conformal symmetry}.
The original Lagrangian function was introduced by Marnelius \cite{Ma-79}
and, subsequently, Siegel used it for describing the behaviour
of these kind of particles \cite{Si-88}.
Recently, Gr\`acia and Roca \cite{GR-93} have carefully studied
the gauge transformations for this system.

The configuration space of this system is $Q=\Real^{d+2}\times\Real$
and is locally coordinated by the set $(q^a,\lambda )$
($a=0,1,\ldots,d+1$), where $\lambda$ is an unphysical
parameter which is introduced in order to make
the description of the system covariant, and it is responsible for
the local scale invariance.
At the Lagrangian level, the system is dynamically described by the
Lagrangian function
$$
\Lag :=\frac{1}{2}g_{ab}(v^a v^b-\lambda q^a q^b )
\in\Cinfty (\Tan Q)
$$
where $g$ is an indefinite metric in $\Real^{d+2}$
with signature ${\rm sign}\, (g_{ab})=(1,-1,\ldots ,-1,1)$. From
here and using the canonical structures of the tangent bundle
$\Tan Q$, we construct the
{\sl Lagrangian 2-form} and the {\sl energy Lagrangian function}:
$$
\omega_{\Lag}=g_{ab}\d q^a\wedge\d v^b
\quad ;\quad
{\rm E}_{\Lag}=\frac{1}{2}g_{ab}(v^a v^b+\lambda q^a q^b)
$$
($v^a$ denote the generalized velocities corresponding
to the coordinates $q^a$). The system is singular
since the generalized velocity $u$ corresponding to the generalized
coordinate $\lambda$ does not appear explicitly in the Lagrangian function.
Hence $\omega_{\Lag}$ is presymplectic and
\dst\ker\,\omega_{\Lag}\equiv
\left\{\derpar{}{\lambda},\derpar{}{u}\right\}\) .
So $(\Tan Q,\omega_{\Lag},{\rm E}_{\Lag})$ is a presymplectic
dynamical system which is not compatible since
\dst\inn\left(\derpar{}{\lambda}\right)\d{\rm E}_{\Lag}\not= 0\) .
Using some of the known stabilization algorithms,
we find that the final constraint submanifold
$j_M\colon M\hookrightarrow\Tan Q$ is defined by the constraints
\cite{GR-93}
$$
\eta_1=\frac{1}{2}g_{ab}q^aq^b \ ,\
\eta_2=g_{ab}v^aq^b \ ,\
\eta_3=g_{ab}v^av^b-\lambda g_{ab}q^aq^b
$$
In this case we can take a basis of constraints
made of presymplectic Hamiltonian functions, for instance
\beann
f_{\xi_1}=\frac{1}{2}g_{ab}q^aq^b \quad ,\quad
f_{\xi_2}=g_{ab}v^aq^b \quad ,\quad
f_{\xi_3}=\frac{1}{2}g_{ab}v^av^b
\eeann
The vector fields which are solutions of the dynamical equation
$$
(\inn (X)\omega_{\Lag}-\d{\rm E}_{\Lag})\vert_M=0
$$
are the following
$$
X\vert_M=
v^a\derpar{}{q^a}+\lambda q^a\derpar{}{v^a}+u\derpar{}{\lambda}+f\derpar{}{u}
$$
(where $f$ is an arbitrary function).

The compatible presymplectic system
$(M,\Omega,{\rm H})=(M,j_M^*\omega_{\Lag},j_M^*{\rm E}_{\Lag})$
exhibits point gauge symmetries
which are infinitesimally generated by the following fundamental vector fields
$$
(\tilde\xi_1,\tilde\xi_2,\tilde\xi_3,\tilde\xi_4,\tilde\xi_5)=
\left( q^a\derpar{}{v^a},v^a\derpar{}{v^a}-q^a\derpar{}{q^a},
v^a\derpar{}{q^a},\derpar{}{\lambda},\derpar{}{u}\right)
$$
Observe that $\tilde\xi_1,\tilde\xi_2,\tilde\xi_3$
are the presymplectic Hamiltonian vector fields corresponding to
$f_{\xi_1},f_{\xi_2},f_{\xi_3}$ respectively
and $\tilde\xi_4,\tilde\xi_5\in\ker\,\Omega$.
They are all tangent to $M$ and, hence,
they make a local basis of $\underline{\ker\,j_M^*\Omega}$.

It is interesting to note that the
system is also invariant under rigid ${\bf O}(2,d)$ rotations.
Nevertheless, it can be shown that there exist
${\bf O}(2,d)$ Lagrangian gauge transformations
(see \cite{GR-93} and \cite{Ma-79}) and hence, in this case,
this group of symmetries is a closed subgroup of the gauge group ${\cal G}$.

Taking all of this into account, the action of ${\cal G}$ on
$(M,j_M^*\omega_{\Lag})$ is strongly presymplectic
(it is in fact an exact action in relation to the {\sl Lagrangian 1-form}
$j_M^*\theta_{\Lag}=j_M^*(g_{ab}v^a\d q^b)$).
Thus, a momentum map ${\cal J}$ can be defined for this action
such that $M={\cal J}^{-1}(0)$. Therefore,
the presymplectic reduction procedure
is simply the well-known gauge reduction for the
compatible presymplectic system
$(M,j_M^*\omega_{\Lag},j_M^*{\rm E}_{\Lag})$.

\section{Conclusions and outlook}

We have made a study about actions of Lie groups on
presymplectic manifolds and the subsequent reduction procedure.
The main results and considerations here discussed are the following:
\bit
\item
We have made the natural extension of the concepts of the
theory of symplectic actions of Lie groups on
symplectic manifolds to this case.
\item
The existence of comomentum and momentum maps
are analyzed, obtaining an obstruction similar to
the symplectic case (but involving
the set $B_h^1(M)/Z_h^1(M)$ instead of the first cohomology
group $H^1(M)$).
\item
We have investigated the properties and characteristics of the
level sets of the momentum map for weakly regular values,
as a standpoint for reduction.
As a particular result, the interpretation of these level sets
as the maximal integral submanifolds of a Pfaff system
allows us to simplify the proof of some results.
We hope that this interpretation will be
of interest with a view to extending the reduction
procedure to field theories.
\item
The reduction of presymplectic manifolds by
presymplectic actions of Lie groups has been achieved
for weakly regular values of the momentum map, following
the guidelines of the symplectic reduction theory of Marsden-Weinstein.
With the usual hypothesis, the reduced phase space is endowed with
a structure of presymplectic manifold, in general.
\item
The concept of symmetry for presymplectic dynamical systems is displayed.
The reduction of compatible and non-compatible
presymplectic dynamical systems with symmetry
is made as an application of the theory just developed.
These results hold both for presymplectic Lagrangian or Hamiltonian systems.

When gauge symmetries are taken together with the non-gauge symmetries
of the system, then the reduced phase space is endowed with
a structure of symplectic manifold with dynamics of Hamiltonian type.
\item
The procedure of {\sl complete presymplectic reduction}
allows us to reach the orbit space in a straightforward way,
in comparison with other step-by-step reduction procedures, namely,
{\sl coisotropic imbedding} plus {\sl symplectic reduction} and
{\sl gauge reduction} plus {\sl symplectic reduction},
which lead to the same final reduced phase space.
The equivalence of all these methods is also proved.
\item
As an example, we have considered non-autonomous dynamical systems.
Starting from the presymplectic formulation of these systems
(which allow us to include also the singular case in a natural way),
we have adapted the notion of symmetry, and then by applying the reduction
procedure previously studied, results similar to those of
other works that have analyzed this problem have been obtained.
The main advantage of the formalism is that the treatment
of the singular case is absolutely ``on way''.
\item
As a particular case, the reduction of time-dependent regular dynamical systems
is considered in the framework of time-invariant non-autonomous systems.
In this case, the reduced phase space is a contact manifold
since the level sets of the momentum map are the energy constant hypersurfaces
and reduction removes the time coordinate from the initial system.
In this way, in our opinion, this is a better result
than those obtained applying the
symplectic reduction techniques to the extended phase space of the system,
since reduction then leads to a symplectic system in the reduced phase space,
but does not directly give the dynamics on the constant-energy hypersurfaces.
\item
Another interesting example is the complete reduction of a
particular case of the {\it Capri-Kobayashi} mechanical
model for field theories coupled to external fields,
exhibiting both gauge and non-gauge symmetries, in the Lagrangian
formalism. It is shown that, under suitable circumstances,
the existence of Lagrangian constraints arising from the search for dynamical solutions which are second
order differential equations is irrelevant in the reduction procedure.
\item
Finally, we have also checked this method by applying it to
a discussion of the gauge reduction of the conformal particle
(in the Lagrangian formalism).
\eit

\appendix

\section{Linear reduction}

In this appendix we wish carry out a quick review of the
reduction theory, giving at the same time a linear algebraic
interpretation of this theory for the general case of linear forms of arbitrary order.

Let $E$ be a linear vector space, with ${\rm dim}\, E=n$, and
a linear form $\alpha\in\Lambda^kE^*$, with $k\geq 2$.
Let $S$ be a subspace of $E$. Then take
$$
S^{\bot_1}:=\{ u\in E\ \vert\ \inn (u)\inn (v)\alpha =0
\ ,\ \forall v\in S\} \equiv N
$$
let $j\colon N\hookrightarrow E$ be the natural inclusion
and $\alpha_N:=j^*\alpha$.

If $v\in N\cap S$, then $\inn (u)\inn (v)\alpha =0$,
for every $u\in N$, and therefore $v\in\ker\,\alpha_N$;
that is, $N\cap S\subset\ker\alpha_N\subset N$.
Then we have the projections
$$
N\mapping{\pi_1} N/N\cap S\mapping{\pi_2}
N/\ker\,\alpha_N =(N/N\cap S)/(\ker\,\alpha_N /N\cap S)
$$
and there exist $\alpha_1\in\Lambda^k(N/N\cap S)$ and
$\alpha_3\in\Lambda^k(N/\ker\,\alpha_N )$ such that
$\alpha_N=\pi_1^*\alpha_1$ and $\alpha_N=\pi_3^*\alpha_3$,
where $\pi_3=\pi_2\circ\pi_1$.

Notice that $\ker\,\alpha_3=\{ 0\}$, because the space $S$, projected
by $\pi_3$, ``has been removed''.
This is a ``reduction'' procedure in the sense that
a subspace is removed from a vector space by a reduction of the dimension.
Note that it is not useful to make the quotient $E/S$ and
then the projection $E\to E/S$ because the form
$\alpha$ does not project onto the quotient unless $S\subset\ker\,\alpha$.
Then, $N=S^{\bot_1}$ is a subspace of $E$ which can be
reduced in such a way that the form $\alpha$ goes down to the quotient.

As a particular situation, we can study the case $k=2$.
Then we can prove that
$$
\ker\,\alpha_N =\ker\,\alpha +N\cap S
$$
In fact; let $\{ e_1,\ldots ,e_k\}$ be a
basis of $S$. If $v\in\ker\alpha_N$, then
$\inn (u)\inn (v)\alpha =0$, for every $u\in N$,
and $N\subset\ker\,\inn (v)\alpha$. But
\dst N=\cap_{j=1,\ldots ,k}\ker\,\inn (e_j)\alpha\) ,
then we have that $\inn (v)\alpha$ is a linear combination of
$\inn (e_1)\alpha ,\ldots ,\inn (e_k)\alpha$ and, therefore,
$v\in (\ker\,\alpha +S)\cap N$. But, since $\ker\,\alpha\subset N$,
the result follows.

If in addition, $\alpha$ is a symplectic linear form;
that is, $\ker\,\alpha =\{ 0\}$, then
$\ker\,\alpha_N=N\cap S$ and we have the unique projection
$$
N\mapping{\pi} N/N\cap S
$$
and a unique form $\hat\alpha\in\Lambda^2(N/N\cap S)$
with $\pi^*\hat\alpha =\alpha$, which is also a symplectic form.
This is the result of the Marsden-Weinstein reduction procedure
in the linear case.

\subsection*{Acknowledgments}

We thank Dr. Xavier Gr\`acia-Sabat\'e (U.P.C.) for
bringing the example of the conformal particle
to our attention and for explaining to us some of its characteristics.
We are greateful for the assistance of the referee, whose suggestions
have enabled us to improve the final version of the work.
We also thank Mr. Jeff Palmer for his assistance in
preparing the English version of the manuscript.

We want also to thank the financial support
of the CICYT TAP97-0969-C03-01.

\end{document}